\input harvmac.tex

\input epsf.tex
\def\figin{\epsfcheck\figin}\def\figins{\epsfcheck\figins}
\def\epsfcheck{\ifx\epsfbox\UnDeFiSIeD
\message{(NO epsf.tex, FIGURES WILL BE IGNORED)}
\gdef\figin##1{\vskip2in}\gdef\figins##1{\hskip.5in}
\else\message{(FIGURES WILL BE INCLUDED)}%
\gdef\figin##1{##1}\gdef\figins##1{##1}\fi}
\def\DefWarn#1{}
\def\figinsert{\goodbreak\midinsert}
\def\ifig#1#2#3{\DefWarn#1\xdef#1{fig.~\the\figno}
\writedef{#1\leftbracket fig.\noexpand~\the\figno}%
\figinsert\figin{\centerline{#3}}\medskip\centerline{\vbox{\baselineskip12pt
\advance\hsize by -1truein\noindent\footnotefont{\bf Fig.~\the\figno:} #2}}
\bigskip\endinsert\global\advance\figno by1}


\def\half{{1 \over 2}}

\lref\rango{S.~Corley, A.~Jevicki and S.~Ramgoolam,
``Exact correlators of giant gravitons from dual N = 4 SYM theory,''
arXiv:hep-th/0111222.
}

\lref\metsaev{
R.~R.~Metsaev,
plane wave Ramond-Ramond  background,''
        arXiv:hep-th/0112044.
}

\lref\giant{
J.~McGreevy, L.~Susskind and N.~Toumbas,
``Invasion of the giant gravitons from anti-de Sitter space,''
JHEP {\bf 0006}, 008 (2000)
[arXiv:hep-th/0003075].
}

\lref\thooft{
G.~'t Hooft,
``A Planar Diagram Theory For Strong Interactions,''
Nucl.\ Phys.\ B {\bf 72}, 461 (1974).
}

\lref\polyakov{
A.~M.~Polyakov,
``String theory and quark confinement,''
Nucl.\ Phys.\ Proc.\ Suppl.\  {\bf 68}, 1 (1998)
[arXiv:hep-th/9711002].
}

\lref\review{
O.~Aharony, S.~S.~Gubser, J.~Maldacena, H.~Ooguri and Y.~Oz,
``Large N field theories, string theory and gravity,''
Phys.\ Rept.\  {\bf 323}, 183 (2000)
[arXiv:hep-th/9905111].
}

\lref\gkp{
S.~S.~Gubser, I.~R.~Klebanov and A.~M.~Polyakov,
``Gauge theory correlators from non-critical string theory,''
Phys.\ Lett.\ B {\bf 428}, 105 (1998)
[arXiv:hep-th/9802109].
}

\lref\jm{
J.~Maldacena,
``The large $N$ limit of superconformal field theories and supergravity,''
Adv.\ Theor.\ Math.\ Phys.\  {\bf 2}, 231 (1998)
[Int.\ J.\ Theor.\ Phys.\  {\bf 38}, 1113 (1998)]
[arXiv:hep-th/9711200].
}

\lref\wittenhol{
E.~Witten,
``Anti-de Sitter space and holography,''
Adv.\ Theor.\ Math.\ Phys.\  {\bf 2}, 253 (1998)
[arXiv:hep-th/9802150].
}

\lref\polyakov{
A.~M.~Polyakov,
``Gauge fields and space-time,''
arXiv:hep-th/0110196.
}

\lref\vijay{
V.~Balasubramanian, M.~Berkooz, A.~Naqvi and M.~J.~Strassler,
``Giant gravitons in conformal field theory,''
arXiv:hep-th/0107119.
}

\lref\longstrings{
N.~Seiberg and E.~Witten,
``The D1/D5 system and singular CFT,''
JHEP {\bf 9904}, 017 (1999)
[arXiv:hep-th/9903224].
 ;J.~Maldacena and H.~Ooguri,
``Strings in AdS(3) and SL(2,R) WZW model. I,''
J.\ Math.\ Phys.\  {\bf 42}, 2929 (2001)
[arXiv:hep-th/0001053].
}

\lref\bfss{T.~Banks, W.~Fischler, S.~H.~Shenker and L.~Susskind,
``M theory as a matrix model: A conjecture,''
Phys.\ Rev.\ D {\bf 55}, 5112 (1997)
[arXiv:hep-th/9610043].
}
\lref\susskind{L.~Susskind,
``Another conjecture about M(atrix) theory,''
arXiv:hep-th/9704080.
}
\lref\sen{A.~Sen,
``D0 branes on T(n) and matrix theory,''
Adv.\ Theor.\ Math.\ Phys.\  {\bf 2}, 51 (1998)
[arXiv:hep-th/9709220].
}
\lref\seiberg{N.~Seiberg,
``Why is the matrix model correct?,''
Phys.\ Rev.\ Lett.\  {\bf 79}, 3577 (1997)
[arXiv:hep-th/9710009].
}

\lref\bsei{
T.~Banks and N.~Seiberg,
``Strings from matrices,''
Nucl.\ Phys.\ B {\bf 497}, 41 (1997)
[arXiv:hep-th/9702187].
}

\lref\fp{
J.~Figueroa-O'Farrill and G.~Papadopoulos,
``Homogeneous fluxes, branes and a maximally 
supersymmetric solution of  M-theory,''
JHEP {\bf 0108}, 036 (2001)
[arXiv:hep-th/0105308].
}
\lref\figueroaiib{
M.~Blau, J.~Figueroa-O'Farrill, C.~Hull and G.~Papadopoulos,
``A new maximally supersymmetric background of IIB superstring theory,''
JHEP {\bf 0201}, 047 (2001)
[arXiv:hep-th/0110242].
}

\lref\figueroarecent{
M.~Blau, J.~Figueroa-O'Farrill, C.~Hull and G.~Papadopoulos,
``Penrose limits and maximal supersymmetry,''
arXiv:hep-th/0201081.
}

\lref\penrose{R. Penrose, ``Any spacetime has a plane wave as a limit'',
Differential geometry and relativity, Reidel, Dordrecht, 1976, pp. 
271-275.
}

\lref\kg{J.~Kowalski-Glikman,
``Vacuum States In Supersymmetric Kaluza-Klein Theory,''
Phys.\ Lett.\ B {\bf 134}, 194 (1984).
}

\lref\planewaves{See for example: 
D.~Amati and C.~Klimcik,
``Strings In A Shock Wave Background And Generation Of Curved Geometry From Flat Space String Theory,''
Phys.\ Lett.\ B {\bf 210}, 92 (1988).
G.~T.~Horowitz and A.~R.~Steif,
``Space-Time Singularities In String Theory,''
Phys.\ Rev.\ Lett.\  {\bf 64}, 260 (1990).
H.~J.~de Vega and N.~Sanchez,
``Quantum String Propagation Through Gravitational Shock Waves,''
Phys.\ Lett.\ B {\bf 244}, 215 (1990);
``Space-Time Singularities 
In String Theory And String Propagation 
Through Gravitational Shock Waves,''
Phys.\ Rev.\ Lett.\  {\bf 65}, 1517 (1990).
O.~Jofre and C.~Nunez,
``Strings In Plane Wave Backgrounds Reexamined,''
Phys.\ Rev.\ D {\bf 50}, 5232 (1994)
[arXiv:hep-th/9311187].
}

\lref\polchinski{
J.~Polchinski,
``String Theory''
{\it  Cambridge, UK: Univ. Pr. (1998)}.
}

\lref\ghmnt{ M.~T.~Grisaru, P.~Howe, L.~Mezincescu, B.~E.~W.~Nilsson,
P.~K.~Townsend, ``N=2 superstrings in a supergravity background,''
Phys.\ Lett.\ B {\bf 162}, 116 (1985) 
}

\lref\wpps{ B.~de~Wit, K.~Peeters, J.~Plefka, A.~Sevrin, ``The 
M-theory two-brane in $AdS_4 \times S_7$ and $AdS_7 \times S_4$,''
Phys.\ Lett.\ B {\bf 443}, 153 (1998)
[arXiv:hep-th/9808052]
}

\lref\krr{ R.~Kallosh, J.~Rahmfeld, A.~Rajaraman, ``Near-horizon 
superspace,'' JHEP {\bf 9809}, 002 (1998) 
[arXiv:hep-th/9805217]
}

\lref\myers{ R. Myers, ``Dielectric-branes'',
JHEP {\bf 12}(1999) 022, hep-th/9910053
}

\lref\kt{ R.~Kallosh, A.~A.~Tseytlin, ``Simplifying the 
superstring action on $AdS_5 \times S_5$,'' JHEP {\bf 9810}, 016 (1998)
[arXiv:hep-th/9808088]
}

\lref\mt{ R.~R.~Metsaev, A.~A.~Tseytlin, ``Superstring action 
in $AdS_5 \times S_5$: k symmetry light cone gauge,''
Phys.\ Rev.\ D {\bf 63}, 046002 (2001)
[arXiv:hep-th/0007036]
}

\lref\mtt{ R.~R.~Metsaev, C.~B.~Thorn, A.~A.~Tseytlin, ``Light cone 
superstring in AdS spacetime,'' Nucl.\ Phys.\ B {\bf 596}, 151 (2001)
[arXiv:hep-th/0009171]
}

\lref\dkss{ S.~Deger, A.~Kaya, E.~Sezgin, P.~Sundell, ``Spectrum of D=6,
N=4b supergravity on $AdS_3 \times S_3$,'' Nucl.\ Phys.\ B {\bf 536},
110 (1998)
[arXiv:hep-th/9804166]
}

\lref\meessen{ P.~Meessen, ``A small note on pp-vacua in 6 and
 5 dimensions,'' 
arXiv:hep-th/0111031.
}

\lref\ps{J.~Polchinski and M.~J.~Strassler,
``The string dual of a confining four-dimensional gauge theory,''
arXiv:hep-th/0003136.
}
\lref\motl{
L.~Motl,
``Proposals on nonperturbative superstring interactions,''
arXiv:hep-th/9701025.
}
\lref\dvv{
R.~Dijkgraaf, E.~Verlinde and H.~Verlinde,
``Matrix string theory,''
Nucl.\ Phys.\ B {\bf 500}, 43 (1997)
[arXiv:hep-th/9703030].
}

\lref\sfetsos{K.~Sfetsos,
``Gauging a nonsemisimple WZW model,''
Phys.\ Lett.\ B {\bf 324}, 335 (1994)
[arXiv:hep-th/9311010].
K.~Sfetsos and A.~A.~Tseytlin,
``Four-dimensional plane wave string solutions with coset CFT description,''
Nucl.\ Phys.\ B {\bf 427}, 245 (1994)
[arXiv:hep-th/9404063].
}

\lref\guven{R.~Gueven,
``Plane Waves In Effective Field Theories Of Superstrings,''
Phys.\ Lett.\ B {\bf 191}, 275 (1987).
 R. Guven,  ``Plane Wave Limits and T-duality'', Phys. Lett. B482(2000)
      255.}


{\Title{\vbox{
\hbox{\tt hep-th/0202021}}}
{\vbox{
\centerline{Strings in flat space and pp waves from }
\vskip .1in \centerline{ ${\cal N}=4$ Super Yang Mills }}}
\vskip .3in
\centerline{David Berenstein, Juan Maldacena and Horatiu Nastase}
\vskip .4in
}

\vskip .1in
\centerline{ Institute for Advanced Study, Princeton, NJ 08540}

\vskip .4in

We explain how the string spectrum in flat space and pp-waves 
arises from the large $N$ limit, at fixed $g^2_{YM}$, of
U(N) ${\cal N} =4$  super Yang Mills. We reproduce the 
spectrum by summing a subset  of the planar Feynman diagrams. We 
give a heuristic argument for why we can neglect other diagrams. 

We also discuss some other aspects of pp-waves and we present
a  matrix
model associated to the DLCQ description of the maximally 
supersymmetric eleven dimensional pp-waves.

\vfill
\eject

\newsec{Introduction}

The fact that large $N$ gauge theories have  a string theory 
description was believed for a long time \thooft .
These strings live in more than four dimensions \polyakov .
One of the surprising aspects of the  AdS/CFT correspondence
\refs{\jm,\gkp,\wittenhol,\review} is 
the fact that for   ${\cal N} =4$  super Yang Mills these strings
move in ten dimensions and are the usual strings of type IIB string
theory. The radius of curvature of the ten dimensional space
goes as $R/l_s \sim (g^2_{YM} N)^{1/4}$. The spectrum of 
strings on $AdS_5 \times S^5$  corresponds to the spectrum of 
single trace operators in the Yang Mills  theory. The perturbative 
string spectrum 
is not known exactly for general values of the 't Hooft coupling, but
it is certainly known for large values of the 't Hooft coupling where 
we have the string spectrum in flat space. In this paper we will explain
how to reproduce this spectrum from the gauge theory point of view. 
In fact we will be able to do slightly better than reproducing 
the flat space spectrum. We will reproduce the spectrum on 
a pp-wave. These  pp-waves incorporate, in a precise sense,  the 
first correction to the flat space result for certain states. 

The basic idea is the following. We consider chiral primary 
operators such as $Tr[Z^J]$ with large $J$. This state corresponds
to a graviton with large momentum $p^+$. Then we consider
replacing some of the $Z$s in this operator by other fields, such
as $\phi$, one of the other transverse scalars. 
The position of $\phi$ inside the operator
will matter since we are in the planar limit. When we include
interactions $\phi$ can start shifting position inside the 
operator. This motion of $\phi$ among the $Z$s is 
described by a field in 1+1 dimensions. We then identify this
field with the field corresponding to one of the transverse 
scalars of a string in light cone gauge. 
This can be shown by summing a subset of the Yang Mills Feynman
diagrams. We will present a heuristic argument for why other diagrams
are not important.

Since these results amount to a ``derivation'' of the string spectrum 
at large 't Hooft coupling from the gauge theory, it is quite plausible
that by thinking along the lines sketched in this paper one could
find the string theory for other cases, most interestingly cases
where the string dual is not known (such as pure non-supersymmetric
Yang Mills).  

We will also describe other aspects of the physics of plane waves.
For example we consider the M-theory plane wave background with 
maximal supersymmetry \refs{\kg,\fp} and we show that there is 
an interesting matrix model describing its DLCQ compactification. 
This matrix model has some unusual features such as the absence
of flat directions. We merely touch the surface on this topic 
in section 5, postponing a more detailed investigation for the 
future.

This paper is organized as follows. 
In section two we will describe a limit of $AdS_5\times S^5$ that
gives a plane wave. 
In section three we describe the spectrum of string theory
on a plane wave. 
In section 4 we describe the computation of the spectrum from
the ${\cal N} =4$ Yang Mills point of view. 
In section 5 we describe the Matrix model associated to the 
DLCQ compactification of the M-theory plane wave and 
discuss some of its features. 
In appendix A we describe in detail some of the computations
necessary for section 4. In appendix B we prove  the supersymmetry
of the Matrix model of section 5. 
In appendix C we describe the string spectrum on a plane wave
with mixed NS and RR backgrounds.

\newsec{pp waves as limits of $AdS \times S$}

In this section we show how pp wave geometries arise as a limit
of $AdS_p \times S^q $ \foot{
While this paper was being written the paper \figueroarecent\ appeared
which contains the same point as this section.}. 
Let us first consider the case of $AdS_5 \times S^5$. 
The idea is to consider the trajectory of a particle that is moving
very fast along the $S^5$ and to focus on the geometry that this 
particle sees. 
We start with the $AdS_5 \times S^5$ metric written as 
\eqn\metric{
ds^2 = R^2 
\left[ -dt^2 \cosh^2\rho + d\rho^2 + \sinh^2\rho d\Omega_3^2 +
  d\psi^2 \cos^2 \theta  + d\theta^2 + \sin^2 \theta d {\Omega'}^2_3
\right]
}
We want to consider a particle moving along the $\psi$ direction 
and sitting at $\rho=0$ and $\theta=0$. We will focus on the 
geometry near this trajectory. We can do this  systematically
by introducing coordinates $\tilde x^\pm = {t \pm \psi \over 2}$ and then 
performing the rescaling
\eqn\rescalings{
x^+ =  \tilde x^+ ~,~~~~~x^- = R^2 \tilde x^- ~,~~~~~~
\rho = { r \over R} ~,~~~~~~  \theta = { y \over R} ~,~~~~ R \to \infty
}
In this limit the metric \metric\ becomes 
\eqn\metricfin{
ds^2 = - 4 dx^+ dx^- -  ( {\vec r}^{\ 2} + {\vec y}^{\ 2}) (dx^+)^2 
+ d {\vec y}^{ \ 2} + d{\vec r}^{\  2}
}
where $\vec y$ and $\vec r$ parametrize points on $R^4$.
We can also see that only the components of $F$ with a plus index 
survive the limit. 
We see that this metric is of the form of a plane wave metric\foot{
The constant in front of $F$ depends on the normalizations of $F$
and can be computed once a normalization is chosen. } 
\eqn\metricgen{\eqalign{
ds^2 =& -4 d x^+ d x^-  - \mu^2  \vec z^{ \ 2} {d x^{+}}^2 + 
d \vec z^{ \ 2}
\cr
F_{+1234} & = F_{+5678} = {\rm const} \times \mu 
}}
where $\vec z$ parametrizes a point in $R^8$. 
The mass parameter $\mu$ can be introduced by rescaling \rescalings\ 
$x^- \to x^-/\mu $ and $x^+ \to  \mu  x^+ $. These solutions where
studied in \figueroaiib .

It will be convenient for us to understand how the energy and 
angular momentum along $\psi$ scale  in the limit \rescalings .
The energy in global coordinates in $AdS$ 
is given by $E = i \partial_t$ and the 
angular momentum by $J = - i \partial_\psi$. This angular 
momentum generator can be thought of as the generator
that rotates the 12 plane of  $R^6$. 
In terms of the dual CFT these are the energy and R-charge
of a state of the field theory on $S^3 \times R$ where 
the $S^3$ has unit radius. Alternatively, we can say 
that $ E= \Delta$ is the conformal dimension of an 
operator on $R^4$. 
We find that 
\eqn\generators{\eqalign{
 2 p^- = - p_+ =& i \partial_{x^+} =i \partial_{\tilde x^+} 
=  i ( \partial_t + \partial_\psi) =  \Delta - J
\cr 
2 p^+ = -p_- =& - {\tilde p_- \over R^2} =
 { 1 \over R^2} i \partial_{\tilde x^-}
= { 1 \over R^2} i ( \partial_t - \partial_\psi) =
{\Delta +J \over  R^2}
}}

Notice that $p^{\pm }$ are non-negative due to the BPS condition
$\Delta \geq |J|$. 
Configurations with fixed non zero 
$p^+$ 
in the limit \rescalings\   correspond to states in $AdS$ 
 with  large angular momentum  $J \sim R^2 \sim N^{1/2}$. 
When we perform the rescalings 
\rescalings\ we take the  $N\to \infty$ limit keeping
the string coupling $g$ fixed  
and we focus on  operators  with $J \sim N^{1/2}$  and 
 $\Delta-J$  fixed.

{}From this point of view it is clear that the full supersymmetry 
algebra of 
the metric \metric\ is a contraction of that of $AdS_5 \times S^5$
\figueroaiib .
This algebra implies that $p^\pm \geq 0$. 


This limit is a particular case of Penrose's limit \penrose \foot{
We thank G. Horowitz for suggesting that plane waves could be obtained
this way.}, see also \refs{\sfetsos,\guven}.
In other $AdS_d \times S^p$ geometries we can take similar limits. 
The only minor difference as compared to the above computation 
is that in general the radius of $AdS_d$ and the sphere are not
the same. Performing the limit for $AdS_7 \times S^4$ or
$AdS_4 \times S^7$ we get the same geometry, the 
maximally supersymmetric plane wave metric discussed in \refs{\kg,\fp}.
For the $AdS_3 \times S^3$ geometries that arise in the
D1-D5 system the two radii are equal and the computation is
identical  to the one we did above for  $AdS_5 \times S^5$. 

In general the geometry could depend on other parameters besides
the radius parameter $R$. It is clear that in such cases 
we could also define other interesting limits by rescaling
these other parameters as well. For example one could consider
the geometry that arises by considering 
D3 branes on $A_{k-1}$ singularities \ref\DouglasSW{
M.~R.~Douglas and G.~W.~Moore,
``D-branes, Quivers, and ALE Instantons,''
arXiv:hep-th/9603167.
}. These correspond to geometries
of the form $AdS_5 \times S^5/Z_k$ \ref\KachruYS{
S.~Kachru and E.~Silverstein,
``4d conformal theories and strings on orbifolds,''
Phys.\ Rev.\ Lett.\  {\bf 80}, 4855 (1998)
[arXiv:hep-th/9802183].
}. 
The $Z_k$ quotient leaves an
$S^1$ fixed in the $S^5$ if we parametrize this $S^1$ by the 
$\psi$ direction and we perform the above scaling limit we find
the same geometry that we had above except that now $\vec y$ in 
\metricfin\ parametrizes an $A_{k-1}$ singularity. It seems 
possible to deform a bit the singularity and scale the deformation 
parameter with $R$ in such a way to retain a finite deformation
in the limit. We will not study these limits in detail below but
they are of clear physical interest.

\newsec{ Strings on pp-waves}

It has been known for a while that strings on pp-wave 
NS backgrounds are exactly solvable \refs{\planewaves}.
The same is true for pp-waves on RR backgrounds. 
In fact, after we started thinking about this  
the  paper  by Metsaev \metsaev\  came out, 
so we will refer the reader to
it for the details. 
The basic reason that strings on pp-waves are tractable 
is that the action dramatically simplifies in light cone
gauge. 

We start with the metric \metricgen\  
and
we  choose light cone gauge $x^+ = \tau$ where $\tau$ is 
the worldsheet time. Then we see that the action 
for the eight transverse directions becomes just 
the action for eight massive bosons. Similarly the 
coupling to the RR background gives a mass for the 
eight transverse fermions. 

So in light cone gauge we have eight massive bosons and 
fermions. It turns out that 16 of the 32 supesymmetries of the 
background are linearly realized in light cone gauge (just
as in flat space). These sixteen supersymmetries commute
with the light cone hamiltonian and so they imply that the
bosons and fermions have the same mass, see 
\metsaev . 

After the usual 
gauge fixing (see \polchinski , \metsaev )
the light cone action becomes 
\eqn\lcact{
S = {1 \over 2 \pi \alpha'} 
\int dt \int_0^{2 \pi  \alpha' p^+ } d \sigma
\left[ \half \dot z^2 - \half z'^2 - \half \mu^2 z^2 
+ i \bar S ( \not \partial  + \mu I) S \right]
}
where $I = \Gamma^{1234}$ and $S$ is a Majorana spinor on the 
worldsheet and a positive chirality SO(8) spinor 
under rotations in the eight transverse directions. 
We quantize this action by expanding all fields in 
Fourier modes on the circle labeled by $\sigma$. 
For each Fourier mode we get a harmonic oscillator
(bosonic or fermionic depending on the field). 
Then the
light cone Hamiltonian is 
\eqn\spectrum{
2p^- =-p_+ = H_{lc} = \sum_{n=-\infty}^{+\infty} N_n \sqrt{
\mu^2 + {  n^2 \over (  \alpha' p^+)^2}}
}
Here $n$ is the label of the fourier mode, $n>0$ label
left movers and $n<0$ right movers.  $N_n$ denotes the total 
occupation number
of that mode, including bosons and fermions.
Note that the ground state energy of bosonic oscillators is
canceled by that of the fermionic oscillators. 

In addition we have the condition that the total momentum on
the string vanishes
\eqn\momconstraint{
P = \sum_{n=-\infty}^\infty  n N_n  =0
}
Note that for 
$n=0$ we also have harmonic oscillators (as opposed to the
situation in flat space).
When 
only the $n=0$ modes are excited we reproduce the spectrum 
of massless supergravity modes 
 propagating on the plane wave geometry. 
A particle propagating on a plane wave geometry with 
fixed $p^+$ feels as if it was on a gravitational potential well, 
it cannot escape to infinity if its energy, $p^-$, is finite. 
Similarly a massless particle with zero $p^+$ can go to $r =\infty$
and back in finite $x^-$ time (inversely proportional to $\mu$). 
This is reminiscent to what happens for particles in $AdS$. 
In the limit that $\mu$ is very small, or in other words if
\eqn\flatlim{
 \mu \alpha' p^+  \ll 1
}
we recover the flat space spectrum. Indeed we see 
 {} from  \metricfin\ that the metric reduces to the flat space
metric if we set $\mu$ to zero. 

It is also interesting to consider the opposite limit, 
where 
\eqn\curvedlim{
 \mu \alpha'  p^+  \gg 1
}
In this limit all the low lying string oscillator modes have 
almost the same energy. 
This limit \curvedlim\ corresponds to a highly curved 
background with RR fields. In fact we will later see that 
the appearance of a large number of light modes is 
expected from the Yang-Mills theory. 

It is useful to rewrite \spectrum\ in terms of the 
variables that are natural from the $AdS_5 \times S^5$ point
of view. We find that the contribution to 
$\Delta - J = 2 p^- $ of each oscillator is  its frequency
which can be written as 
\eqn\spectrumads{
(\Delta- J)_n =  w_n = \sqrt{1 +  {4 \pi g N n^2 \over J^2} }
}
using \generators\ and the fact that the $AdS$ radius is
given by $R^4 = 4 \pi g N \alpha'^2$. 
Notice that $N/J^2$ remains fixed in the $N\to \infty $ limit
that we are taking.

It is interesting to note that in the plane wave \metricgen\
 we can also 
have giant gravitons as we have in $AdS_5 \times S^5$. These
giants are D3 branes classically sitting at fixed $x^-$ and wrapping 
the $S^3$ of the first four directions or the $S^3$ of the second four 
directions with a size
\eqn\size{
r^2  =  2 \pi g p^+ \mu \alpha'^2 
}
where $p^+$ is the momentum carried by the giant graviton. This
result 
follows in a straightforward fashion from the results in \giant. 
Its $p^-$ eigenvalue is zero.
We see that the description of these states in terms of D-branes
is correct when their size is much bigger than the string scale. 
In terms of the Yang-Mills variables this happens when 
${J^2 \over N }\gg { 1 \over g} $

There are many other interesting 
aspects of perturbative string propagation 
on plane waves that one could study. 
In appendix C we discuss the spectrum of  strings on plane wave 
background of mixed NS and RR type. 
Note that for more general plane waves, 
for which the factor multiplying $(dx^+)^2$ is not quadratic, 
the action in light cone gauge is  a  more general
interacting massive theory. We could have, for example,
a Landau-Ginsburg theory. It would be nice to analyze these
cases in detail. We can also have an $x^+$ dependent function, 
as discussed in \planewaves .

It is well known
that in conformal gauge the equation of motion for the background 
is conformal invariance of the two dimensional worldsheet theory.
It would be nice to understand what  the equation of motion for
the background is in these more general massive cases, where we have chosen
 the light cone gauge fixing instead. 
In flat space  conditions like $D=26$ appear, in light
cone gauge, from the proper realization of the non-linearly 
realized Lorentz generators.  
These plane wave backgrounds  generically break those Lorentz
generators.

\newsec{ Strings from ${\cal N} =4$ Super Yang Mills }

We are interested in the limit  $N \to \infty$ where
$g^2_{YM}$ is kept fixed and small, $g^2_{YM} \ll 1$.
 We want to consider states
which carry parametrically large 
R charge $J \sim \sqrt{N}$.\foot{
 For reasons that we will discuss later
we also need that $J/N^{1/2}  \ll 1/g_{YM} $. This latter condition 
comes from demanding that \size\ is smaller than the string scale 
and it ensures that the states we consider are strings and not D-brane
``giant gravitons'' \giant . }
This R charge generator, $J$, 
is the SO(2) generator  rotating two of the six scalar fields. 
We want to find the spectrum of states with 
$\Delta - J$ finite  in this limit.  We 
are interested in single trace states of the Yang Mills theory on 
$S^3\times R$, or equivalently, the spectrum of dimensions
of single trace operators of the theory on $R^4$. 
We will often go back and forth between the states and the 
corresponding operators.

Let us first start by understanding the operator  with lowest
value of $\Delta - J=0$. There is a unique single trace 
operator with $\Delta -J=0$, namely 
$Tr[Z^J]$, where $Z\equiv  \phi^5 + i \phi^6$ and
the trace is over the $N$ color indices. 
We are taking $J$ to be the SO(2) generator rotating the plane 56.  
At weak coupling the dimension of this operator is $J$ since 
each $Z$ field has dimension one. 
This operator is a chiral primary and hence its  dimension is protected by 
supersymmetry. It is associated to the vacuum state in light cone 
gauge, which is the unique state with zero light cone hamiltonian. 
In other words we have the correspondence 
\eqn\statezero{
{ 1 \over \sqrt{J}  N^{J/2}} Tr[Z^J]   \longleftrightarrow 
|0,p_+ \rangle_{l.c.} 
}
We have normalized the operator as follows. 
When we compute $ \langle Tr[\bar Z^J](x)  Tr[Z^J](0) \rangle$ 
we have $J$ possibilities
for the contraction of the first $\bar Z$ but then planarity 
implies that we contract the second $\bar Z$ with a $Z$ that is next
to the first one we contracted and so on. Each of these contraction
gives a factor of $N$. Normalizing this two point function to
one we get the normalization factor in \statezero .\foot{ 
In general in the free theory any contraction of a single trace 
 operator with
its complex conjugate one will give us a factor of $N^{n}$, 
where $n$ is the number of fields appearing in the operator. 
If the number of fields is very large it is possible that non-planar 
contractions dominate over planar ones \refs{\vijay,\rango} . 
In our case, due
to the way we scale $J$ this does not occur in the free theory. }

Now we can consider other operators that we can build in the free
theory. We can add other fields, or we can add derivatives of 
fields like $\partial_{(i_1 } \cdots \partial_{i_n)}\phi^r$, where
we only take the traceless combinations since the traces can be 
eliminated via the equations of motion. The order in which these
operators are inserted in the trace is important. All operators 
are all ``words'' constructed by these fields up to the cyclic symmetry, 
these were discussed and  counted 
in  \polyakov . 
We will find it convenient to divide all fields, and derivatives of 
fields,  that appear
in the free  theory according to their $\Delta - J$ eigenvalue. 
There is only one mode that 
has $\Delta-J =0$, which is the mode used in \statezero . 
There are eight bosonic  and eight fermionic 
modes with $\Delta - J =1$. They arise as follows. 
First we have the four scalars in the directions not rotated
by $J$, i.e. $\phi^i$, $i=1,2,3,4$. Then we have
derivatives of the field $Z$,  $ D_i Z = \partial_i Z + [A_i,Z]$,
 where $i=1,2,3,4$ are four directions in $R^4$. 
Finally  there are  eight fermionic operators $\chi^a_{J=\half}$ which 
are the eight components with $J= \half$ of the sixteen 
component gaugino $\chi$ (the other eight components have $J=-\half$). 
These eight components transform in the positive chirality spinor
representation of $SO(4)\times SO(4)$ \foot{
The first SO(4) corresponds to rotations in $R^4$, the space where the 
Yang Mills theory is defined, the second $SO(4) \subset SO(6)$ corresponds
to rotations of the first four scalar fields, this is
the subgroup of $SO(6)$ that commutes with the $SO(2)$, generated by $J$, 
 that we singled
out to perform the analysis. By positive chirality in $SO(4)\times SO(4)$
we mean that it has positive chirality under both $SO(4)$s or
negative under both $SO(4)$. Combining the spinor indices into 
$SO(8)$, $SO(4)\times SO(4) \subset SO(8)$ it has positive chirality
under $SO(8)$. Note that $SO(8)$ is not a symmetry of the background.}.
 We will focus first on operators built out
of these fields and then we will discuss what happens when we include
other fields, with $\Delta-J>1$, such as $\bar Z$.

The state \statezero\ describes a particular mode of ten
dimensional  supergravity 
in a particular wavefunction \wittenhol . 
Let us now discuss how to generate 
all other massless supergravity modes.  On the string theory side 
we construct all these states by applying the zero momentum oscillators
$a_0^i$, $i=1, \dots ,8$
and $S^b_0$, $b=1, \dots 8$ on the light cone vacuum $|0,p_+\rangle_{l.c.}$. 
Since the modes on the string are massive all these zero momentum oscillators
are harmonic oscillators, they all have the same light cone energy. 
So the total light cone energy is  equal to the total number of
oscillators that are acting on the light cone ground state. 
We know that in  $AdS_5\times S^5 $  all 
gravity modes are 
in the same supermultiplet
as the state of the form \statezero \ref\GunaydinFK{
M.~Gunaydin and N.~Marcus,
``The Spectrum Of The S**5 Compactification Of The Chiral N=2, D = 10 Supergravity And The Unitary Supermultiplets Of U(2, 2/4),''
Class.\ Quant.\ Grav.\  {\bf 2}, L11 (1985).
H.~J.~Kim, L.~J.~Romans and P.~van Nieuwenhuizen,
``The Mass Spectrum Of Chiral N=2 D = 10 Supergravity On S**5,''
Phys.\ Rev.\ D {\bf 32}, 389 (1985).
}. 
The same is clearly true in the 
limit that we are considering. 
More precisely, the action of all supersymmetries and bosonic 
symmetries of the plane wave background (which are intimately 
related to the $AdS_5\times S^5$ symmetries) generate all 
other ten dimensional massless modes with given $p_+$. 
For example, by acting by some of the rotations of $S^5$ that
do not commute with the SO(2) symmetry that we singled out 
we create states of the form 
\eqn\create{
 {1 \over \sqrt{J} }\sum_{l}  {1 \over \sqrt{J} N^{J/2 +1/2} }
Tr[ Z^l \phi^r Z^{J-l}] =  {1 \over N^{J/2 +1/2} }Tr[\phi^r Z^J]
}
where $\phi^r$, $r=1,2,3,4$  is one of the scalars neutral
under $J$.
In \create\ we used   the cyclicity of the trace.
Note that we have normalized the states appropriately in the 
planar limit.  
We can act any number of times by these generators and we get 
operators  roughly of the form 
$\sum Tr[ \cdots z \phi^r z\cdots z \phi^k ] $. 
where the sum is over all the possible orderings of the $\phi$s. 
We can repeat this discussion with the other $\Delta-J=1$ fields. 
Each time we insert a new operator we sum over all possible 
locations where we can insert it. Here we are neglecting 
possible extra terms that we need when two $\Delta-J=1$ fields
are at the same position, these are subleading in a $1/J$ expansion
and can be neglected in the large $J$ limit that we are considering.
 In other words, when we act with 
the symmetries that do not leave $Z$ invariant we will change 
one of the  $Z$s in \statezero\ to a field with $\Delta-J=1$, 
when we act again with one of the symmetries we can change 
one of the $Z$s that was left unchanged in the first step or 
we can act on the field that was already changed in the first step.
This second possibility is of lower order in a $1/J$ expansion and
we neglect it. We will always work in a ``dilute gas'' approximation
where most  of the fields in the operator are $Z$s and there 
are a few other fields sprinkled in the operator.

For example, a state with two excitations will be of the form 
\eqn\statetwo{
\sim { 1 \over N^{J/2 +1}} 
  {1\over \sqrt{J} } \sum_{l=1}^J  Tr[\phi^r Z^l \psi_{J=\half}^b  
Z^{J-l} ]
}
where we used the cyclicity of the trace to put the $\phi^r$ operator
at the beginning of the expression. 
We associate \statetwo\ to the string state 
$ a_0^{\dagger k} S^{\dagger \ b}_0 |0,p_+\rangle_{l.c.}$. 
Note that for planar diagrams it is very important to keep
track of the position of the operators. For example, two 
operators of the form $ Tr[\phi^1 Z^l \phi^2 Z^{J-l} ]$ with 
different values of $l$ are orthogonal to each other in the planar limit
(in the free theory). 

The conclusion is that there is a precise correspondence between 
the supergravity modes and the operators. This is of course
well known \refs{\gkp,\wittenhol,\review}. 
Indeed, we see from \spectrum\ that
their $\Delta-J = 2p^- $ is indeed what we compute at weak 
coupling, as we expect from the BPS argument.

In order to understand non-supergravity modes in the bulk it is 
 clear that what we need to understand  the Yang Mills 
description of the states obtained by 
the action of the string oscillators which have $n\not = 0$. 
Let us consider first one of the string
oscillators which creates a bosonic mode along one of the 
four directions that came from the $S^5$, let's say 
$a^{\dagger \ 8}_n$. We already understood that the action of 
$a^{\dagger \ 8}_0$ 
corresponds to insertions of an operator $\phi^4$ on all 
possible positions along the ``string of $Z$'s''. By a 
``string of $Z$s'' we just mean a sequence of $Z$ fields 
one next to the other such as we have in \statezero .  
We propose that $a_n^{\dagger 8} $ corresponds to the insertion of
the same field $\phi^4$ but now with a position dependent 
phase 
\eqn\stateop{
  { 1 \over \sqrt{J} } \sum_{l=1}^J  { 1 \over 
\sqrt{J} N^{J/2 +1/2}} Tr[ Z^l  \phi^4 Z^{J-l} ] 
e^{ 2 \pi i n l \over J}
}
In fact the state \stateop\ vanishes by cyclicity of the trace. 
This corresponds to the fact that we have the constraint that 
the total momentum along the string should vanish \momconstraint ,
 so that 
 we cannot insert only one $a_n^{\dagger \ i}$ oscillator.
So we should  
insert more than one  oscillator so that the total momentum 
is zero.  
For example  we can  consider the string state obtained
by acting with the $a^{\dagger \ 8}_n$ and $a^{\dagger \ 7 }_{-n}$, 
which has
zero total momentum along the string. We propose that this state
should be identified with  
\eqn\statetwomom{
a^{\dagger \ 8}_n a^{\dagger \ 7 }_{-n} |0,p_+\rangle_{l.c.} 
\longleftrightarrow 
 { 1 \over \sqrt{J} } \sum_{l=1}^J  {1 \over N^{J/2 +1}}
 Tr[ \phi^3 Z^l  \phi^4 Z^{J-l} ] 
e^{ 2 \pi i n l \over J}
}
where we used the cyclicity of the trace to simplify the expression. 
The general rule is pretty clear, for each oscillator mode along 
the string we associate one of the $\Delta -J =1$ fields of the 
Yang-Mills theory and we sum over the insertion of this field
at all possible positions with a phase proportional to the 
momentum. 
States whose total momentum is not zero along the string
lead to operators that are
automatically zero by cyclicity of the trace. In this way we
enforce the $L_0 - \bar L_0 =0$ constraint \momconstraint\ 
 on the string spectrum. 

In summary, each string oscillator corresponds to the insertion
of a $\Delta -J =1$ field, summing over all positions with an 
$n$ dependent phase, according to the rule
\eqn\summary{\eqalign{
 a^{\dagger i} & \longrightarrow  D_iZ ~~~~{\rm for}~i=1, \cdots, 4
\cr
a^{\dagger j} &  \longrightarrow  \phi^{j-4} ~~~~
{\rm for}~j=5, \cdots, 8
\cr
S^a  &  \longrightarrow  \chi_{J=\half}^a
}}

In order to show that this identification makes sense
we want to compute the conformal dimension, or more
precisely $\Delta - J$,  of these operators at large 
't Hooft coupling and show that it matches  \spectrum . 
First note that if we set ${gN \over J^2 } \sim 0$ in 
\spectrumads\ we find that all modes, independently of $n$ 
have the same energy, namely one. This is what we find at
weak 't Hooft coupling where all operators of the form 
\statetwomom\ have the same energy, independently of $n$. 
Expanding the string theory result \spectrumads\  we find that 
the first correction is of the form 
\eqn\firstcorr{
 (\Delta- J)_n =  w_n = 1 + { 2 \pi g N n^2 \over J^2} + \cdots
}

This looks like a  first order correction in the 't Hooft
coupling and we can wonder if we can reproduce it by a 
 a simple perturbative computation. Manipulations with non BPS
operators suggest that anomalous dimensions grow like
$g^2N$ and that they disappear from the spectrum of the theory at strong
coupling. However, this line of reasoning assumes that we keep the 
dimension of the operator in the free field
theory  ($J$ in this case) fixed as we take the large 
$N$ limit. In our case the states we begin with are almost BPS; there are
cancellations which depend on the free field theory dimension ($J$)
which render the result finite even in the infinite 't Hooft 
coupling limit.
The interesting diagrams arise from the following interaction
vertex
\eqn\interterm{
 \sim g^2_{YM}  Tr( [Z,\phi^j][\bar Z, \phi^j])
}
\ifig\exchange{ Diagrams that exchange the position of $\phi$. They 
have ``momentum'', $n$, dependent
contributions.}
{\epsfxsize 1 in\epsfbox{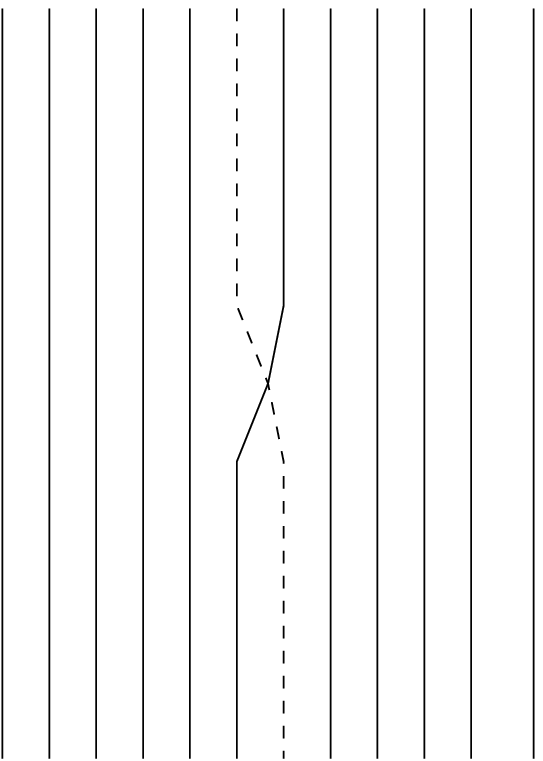}}

This vertex leads to diagrams, such as shown in \exchange\
which move the
position of the $\phi^j$ operator along the ``string'' of $Z$'s. 
In  the free theory, once a $\phi^j$ operator is inserted
at one position along the string it will stay there, states with 
$\phi^j$'s at different positions are orthogonal to each other
in the planar limit (up to the cyclicity of the trace). 
We can think of the string of $Z$s in \statezero\ as defining a
lattice, when we insert an operator $\phi^1$ at different 
positions along the string of $Z$s we are exciting an oscillator  
$b^{\dagger}_l$  at the site $l$ on the lattice, $l=1, \cdots J$.
 The interaction 
term \interterm\ can  take an excitation from one site 
in the lattice to the neighboring site. So we see that the 
effects of  \interterm\ will be sensitive to the 
momentum $n$. 
In fact one can precisely reproduce \firstcorr\ from \interterm\
including the precise numerical coefficient. In appendix A 
we give the  details of this  computation.

Encouraged by the success of this comparison 
we want  to reproduce the full square root\foot{
Square roots of 
the 't Hooft coupling are  ubiquitous in the AdS computations.}
in 
\spectrumads . At first sight this seems a daunting 
computation since it involves  an infinite number of 
corrections. These corrections nevertheless can be 
obtained from exponentiating  \interterm\  and taking 
into account that in \interterm\ there are terms 
involving two creation operators $b^\dagger$ and 
two annihilation operators $ b$. In other words we 
have $\phi \sim b + b^\dagger$. 
As we explained above, we can view $\phi$'s at different
positions as different operators. So we introduce an 
operator $b_l^\dagger$ which introduces a $\phi$ operator
at the site $l$ along the string of $Z$s. 
Then the free hamiltonian plus the 
 interaction term \interterm\ can be thought of 
as 
\eqn\interlattice{
 H \sim  \sum_i  b_l^\dagger b_l +  { g^2_{YM} N \over (2 \pi)^2 } 
[( b_l + b_l^\dagger)  -( b_{l+1} + b_{l+1}^\dagger)]^2
}
In appendix A we give more details on the derivation of 
\interlattice . 
In the large $N$ and $J$ 
 limit it is clear that \interlattice\ reduces
to the continuum Hamiltonian 
\eqn\contver{
H =  \int_0^{L } d\sigma \ \  \half \left[
\dot \phi^2  + \phi'^2 +  \phi^2 \right]~,~~~~~~~
L  = J {\sqrt{\pi \over g N}} \sim p^+
}
which in turn is the correct expression for $H=p^-=\Delta -J$ for
 strings in the light cone gauge.

In summary, the   ``string of $Z$s''  becomes the 
physical string and that each $Z$ carries one unit of 
$J$ which is one unit of $p^+$. Locality along the 
worldsheet of the string comes from the fact that 
planar diagrams  allow  onlycontractions of neighboring 
operators. So the Yang Mills theory gives a string bit
model (see \ref\thorn{
R.~Giles and C.~B.~Thorn,
``A Lattice Approach To String Theory,''
Phys.\ Rev.\ D {\bf 16}, 366 (1977).
C.~B.~Thorn,
``Reformulating string theory with the 1/N expansion,''
arXiv:hep-th/9405069.
})  where each bit is a $Z$ operator. Each bit
carries one unit of $J$ which through \contver\ is  one unit of $p^+$.

The reader might, correctly, be thinking  that all 
this seems too  good  to be true. In fact, we have  neglected
many other diagrams and many other operators which, at 
weak 't Hooft coupling also have small $\Delta-J$. 
In particular, we considered operators which arise by 
inserting the fields  with $\Delta-J=1$ but we did 
not consider the possibility of inserting fields  
corresponding to $\Delta - J=2, 3, \dots$, such as 
$\bar Z, ~ \partial_k \phi^r, ~ \partial_{(l} \partial_{k)} Z$, etc.. 
The diagrams of the type we considered above would give rise
to other 1+1 dimensional fields for each of these modes. 
These are 
present at weak 't Hooft 
coupling but they should not be present at 
 strong coupling, 
since we do not see them  in the string spectrum. 
We believe that what happens is that these fields
get a large mass in the $N\to \infty$ limit. In other
words, the operators get a large conformal dimension. 
In appendix A, we discuss the computation of the first
correction to the energy (the conformal weight) of the 
of the state that results from inserting $\bar Z$ with some 
``momentum'' $n$. In contrast to our previous 
computation for $\Delta-J=1$ fields we find that
besides an effective kinetic term as in \firstcorr\ there is 
an $n$ independent contribution that goes as $gN$ with no
extra powers of $1/J^2$. This is an indication that these
excitations become very massive in the large $gN$ limit. 
In addition, we can compute the decay amplitude of
$\bar Z$ into a pair of $\phi$ insertions. This is also 
very large, of order $gN$. 

Though we have not done a similar  computation for other fields
with $\Delta -J>1$, we believe that the same will be true
for the other fields. 
In general we expect to find many terms in the effective
Lagrangian with  coefficients that are of order
$gN$ with no inverse powers of $J$ to suppress them.
In other words, the  lagrangian of Yang-Mills on $S^3 $ 
acting on a state which contains a large number of $Z$s  
gives  a lagrangian on a discretized spatial circle 
 with an infinite number
of KK modes. The coefficients of this effective lagrangian 
are factors of $gN$, so all fields will generically get very 
large masses. 

The only fields that will not get a large mass are those
whose mass is protected for some reason. The fields
with $\Delta-J=1$ correspond to Goldstone bosons and fermions
of the symmetries broken by the state \statezero . 
Note that despite the fact that they morally are Goldstone
bosons and fermions, their mass is  non-zero, due to the
fact that the symmetries that are broken do not commute with $p^-$, 
the light cone  Hamiltonian. The point is that their masses
are determined, and hence protected, by the (super)symmetry algebra.

Having described how the single string Hilbert space arises it is 
natural to ask whether we can incorporate properly the string 
interactions. 
Clearly string interactions come when we include non-planar diagrams
\thooft . 
There are non-planar diagrams coming from the cubic vertex which 
are proportional to $g_{YM}/N^{1/2}$. These go to zero in the 
large $N$ limit. There are also non-planar 
contributions that come from iterating
the three point vertex or from the quartic vertex in the action.
These are of order $g^2_{YM} \sim g$ compared to planar diagrams
so that we get the right dependence on the string coupling $g$. 
In the discussion in this paragraph we have ignored the fact that
$J$ also becomes large in the limit we are considering. If we naively 
compute the factors of $J$ that would appear we would seem 
to get a divergent contribution for the non-planar diagrams in this
limit. Once we take into account that the cubic and quartic vertices
contain commutators then the powers of $J$ get reduced. 
{}From the gravity side  we expect that some string interactions
should become strong when $ { J\over N^{1/2} } \sim { 1 \over g_{YM} } $.
In other words, at these values of $J$ we expect to find D-brane states
in the gravity side, which means that the usual single trace description
of operators is not valid any more, see discussion around \size .  
We have not been able to successfully reproduce this bound from the 
gauge theory side.

Some of the arguments used in this section look very reminiscent of
the DLCQ description of matrix strings \motl \dvv . It would 
be interesting to see if one can establish a connection between
them.
Notice that the 
DLCQ description of ten dimensional IIB theory is in 
terms of the M2 brane field theory. Since here we are extracting
also a light cone description of IIB string theory we expect that
there should be a direct connection.  

It would also be nice to see if using any of these ideas we can get
a better handle on other large $N$ Yang Mills theories, particularly 
non-supersymmetric ones. The mechanism by which strings appear
in this paper is somewhat reminiscent of \ref\KlebanovBA{
I.~R.~Klebanov and L.~Susskind,
Nucl.\ Phys.\ B {\bf 309}, 175 (1988).
}.

\newsec{ The  matrix model for the DLCQ description 
of M-theory plane waves}

In this section we point out that there is a nice, simple 
matrix model associated to these backgrounds. 
The M-theory pp-wave background is 
\eqn\metricelven{\eqalign{
ds^2 = & -4 dx^-dx^+  - [({\mu \over 3})^2(
 x_1^2 +x_2^2 + x_3^2 ) +({\mu \over 6})^2( x_4^2 + \dots x_9^2)] {dx^+}^2
+ d\vec x^2 
\cr
F_{+123} = &  \mu
}}

This metric arises as a limit similar to the one explained in
section 2 for $AdS_4 \times S^7$ or $AdS_7 \times S^4$
(both cases give the same metric), see also \figueroarecent . 

This metric has a large symmetry group with 32 supersymmetries, 
the algebra is a contraction of the $AdS_{4,7}\times S^{7,4}$
 superalgebras as expected
from the fact that they are limits of the $AdS_{4,7} \times S^{7,4}$
 superalgebras. 
In analogy to the discussion  \refs{\bfss,\susskind,\sen,\seiberg} 
we do DLCQ along the direction $x^- \sim x^- + 2 \pi  R$, and
we consider the sector of the theory with momentum $2 p^+ = - p_- = N/R$. 
Then the dynamics of the theory in this sector is given by 
the  $U(N)$ matrix model
\eqn\action{\eqalign{
S =& S_0 + S_{mass}
\cr
S_0 =&  \int dt Tr\left[
\sum_{j=1}^9  {1 \over 2 (2R)} (D_0  \phi^j)^2  +\Psi ^T D_0\Psi
+{(2R) \over 4 }  \sum_{j,k=1}^9 [\phi^j, \phi^k]^2 +  \right. \cr
& ~~+ \left. \sum_{j=1}^9 
i {(2R) } (\Psi^T \gamma^i [\Psi, \phi^j]) \right]
\cr
S_{mass} =& \int dt  Tr\left[ { 1 \over 2 (2R)} \left(
 -({\mu \over 3})^2 \sum_{j=1,2,3}
(\phi^j)^2 -({\mu \over 6})^2 \sum _{j=4}^9 (\phi^j)^2 \right) 
-{\mu \over 4}\Psi^T \gamma_{123} \Psi \right. \cr
 & ~~~~\left.
- {\mu \over 3}i \sum_{j,k,l=1}^3Tr (\phi^j \phi^k \phi^l) \epsilon_{jkl}
 \right]
}}
where we have set $l_p$=1. We also have that 
$t = x^+$  and  $\phi = {r \over 2\pi}$ 
where $r$ is the physical distance
in eleven dimensions. 
 $S_0$ is the usual matrix theory of \bfss\ \foot{ 
To compare with \bfss\ note 
that due to the form of the metric and the way we define $R$, 
$ 2 R_{our} = R_{BFSS}$. We normalize  $l_p$  
so that $\sqrt{\alpha'} = l_p g^{-1/3}$ when we go to the IIA theory.}. 
 $S_{mass}$ adds mass to the scalar fields and fermion fields, 
plus a term associated to  the Myers effect \myers.

The action \action\ 
 has the transformation rules 
\eqn\susyrules{\eqalign{
\delta \phi^i&=\Psi^T \gamma^i \epsilon (t) \cr
\delta \Psi &= \left( {1 \over (2R)} 
 {D_0 \phi}^i\gamma^i  +{\mu \over 6 (2R) }\sum_{i=1}^3
\phi^i \gamma^i \gamma_{123} -{\mu \over 3 (2R) }\sum_{i=4}^9 \phi^i 
\gamma^i \gamma_{123} + { i \over 2}  [\phi^i, \phi^j] \gamma_{ij} 
\right)\epsilon(t)
\cr
\delta A_0 & = \Psi^T \epsilon(t)
\cr
\epsilon (t) &= e^{-{\mu \over 12} \gamma_{123} t} \epsilon_0
}}
In appendix B we show that the action  \action\ is determined by the
supersymmetry algebra of the plane wave metric \fp.
The matrix model Hamiltonian, associated to this action is equal 
to $H = -p_+$. 

Note that the bosons and fermions have different masses, 
three of the bosons have mass $\mu /3$ while six of them have mass
$\mu /6$.
On the other hand all the fermions have mass $\mu /4$ . 
This is possible because the supersymmetries  \susyrules\ are time
dependent and therefore do not commute with the Hamiltonian. 
This is in agreement with the susy algebra of plane waves \fp , 
see appendix B. 
It is easy to check that the vacuum energy is still zero. 
This is good since there is a state with zero $p_+$ which corresponds
to a single type of graviton mode.

Let us look at the fully supersymmetric solutions  of this action.
Imposing that $\delta \Psi =0$ we find that the only solutions
are
\eqn\fuzzy{
[\phi^i, \phi^j]=i { \mu \over 6 R}
 \epsilon_{ijk} \phi^k \;\;\; i,j,k=1,2,3 \;\;\;\; \dot{\phi}^i=0
\;\; {\rm for\; all \;}i{\rm  \; ~ and ~}\; \phi^i=0, \;\;\; i=4,..,9
}
that is, a fuzzy sphere in the 1,2,3 directions of physical radius 
\eqn\radius{
r \sim 2 \pi \sqrt{ Tr[ \sum_i {\phi^i}^2] \over N } \sim 
 \pi {\mu \over 6 }{N \over R}
}
We see that the mass terms remove completely the moduli space and 
leave only a discrete set of solutions, after modding out by gauge
transformations. This is convenient, as the structure of the ground
states is governed by the semiclassical approximation. One does not need
to solve the full quantum mechanical problem of the ground state 
wave function, an issue which frequently arises in the more standard
matrix model \bfss\  and that has proved very difficult to approach.

The solutions are labelled by all possible  ways 
of dividing an $N$ dimensional representation of SU(2) into
irreducible representations. This number is equal to the number
of partitions of $N$, which is also the number of  multiple 
 graviton 
states with $p_+ =0$, $-p_-={ N \over R}$ in a naive fock space 
description. 

The solutions \fuzzy\ are related to ``giant gravitons'' in the 
plane wave background \metricelven\  which 
are M2 branes wrapping the $S^2$ given 
by $\sum_{i=1}^3 x_i^2 $=constant  and classically sitting at 
a fixed position $x^-$, but with nonzero momentum $p_-$ (but
zero light cone energy $-p_+=0$).
 The supergravity computation of
the radius, similar to that in \giant\  gives again \radius . 

The plane wave geometry also admits giant gravitons which are
M5 branes wrapping the $S^5$ given by $\sum_{i=4}^9 x_i^2=$constant. 
We can similarly compute the value of the radius from the supergravity
side and we get 
\eqn\radiusmfive{
r^4 = { 8 \pi^2 \over 3} \mu  (-p_-)
}
This does not appear as a classical solution of the matrix 
model \action , which is of course not unrelated to the difficulty
of seeing the M5 brane in the matrix model  \ref\BanksNN{
T.~Banks, N.~Seiberg and S.~H.~Shenker,
``Branes from matrices,''
Nucl.\ Phys.\ B {\bf 490}, 91 (1997)
[arXiv:hep-th/9612157].
}. 
It is interesting to notice however that if we write the 
two sphere radius \radius\  in terms of the coupling constant
of the matrix model we find $\hat \phi \sim \mu/g$, where  
 the action with an overall factor of ${1 \over g^2}$. 
This scaling of $\hat \phi$ is precisely what we expect for a classical 
solution. On the other hand, if we express \radiusmfive\ in 
the same way we obtain $ {\hat \phi}^4 \sim 1/g $. 
This scaling with $g$ 
does not correspond to a classical solution of $\action $ and
therefore it is natural that we do not find it.
 The situation seems similar to the one encountered in 
the analysis of vacua of mass deformed ${\cal N}=4$ Yang Mills done
by Polchinski and Strassler \ps . They find that the process 
of D3 branes blowing up into D5 branes can be described classically
in the Yang-Mills theory, while the process of D3 branes 
blowing up into NS 5 branes requires that one takes into account
the quantum effects. 
It is therefore natural to conjecture that the vacuum with 
$x^i =0$ in the quantum mechanics theory corresponds to 
a single large M5 brane.

It is interesting to note that there are other solutions that
preserve a fraction of the supersymmetry and that are time 
dependent. These are commuting configurations of the type
\eqn\halfsusy{\eqalign{
(\phi^4+i\phi^5)(t)&= e^{-i{ \mu \over 6} t } (\phi^4+ i \phi ^5)(0)\cr
&[\phi^i(0), \phi^j(0)]=0, \;\;\; i=4,5\cr
&\gamma_{12345}\epsilon_0=\epsilon_0
}}
and similar ones obtained by 
 replacing 4,5 with any other pair of indices out of 4,..,9, as well
as a similar solution with a pair of indices from $1,2,3$ with exponent
$ e^{ -i{ \mu \over 3} t }$. 

There are many other interesting questions regarding plane waves, such
as the precise nature of the observables, etc. 
They also seem to admit a holographic description, since 
as we remarked above
plane waves have much in common with $AdS$. 
We plan to continue investigating these questions.

{\bf Acknowledgements}

We would like to thank  R. Dijkgraaf, I. Klebanov
N. Seiberg, H. Verlinde and E. Witten
for discussions.

This  research
was supported in part by DOE grants DE-FGO2-91ER40654 and
 DE-FG02-90ER40542.

\appendix{A}{ More detailed computations}

In the first subsection of this 
 appendix we describe in  more detail  the computation of the 
numerical coefficient in \firstcorr .
In the second subsection we discuss how to exponentiate those
corrections to  obtain \spectrumads . 
Finally, in the third  subsection we explain how some $\Delta-J$=2
excitations get a large mass and decay rapidly to $\Delta -J$=1
excitations.

\subsec{ Computation of the first perturbative correction}

In this subsection we discuss the computation of the first
perturbative correction to the anomalous dimension of an 
operator of the form \statetwomom . 
To compute  we analyze  
the correlation function of two such operators. 
We consider operators containing a large number, $J$, of $Z$s with
a few $\phi$s distributed along the ``string of $Z$s''. 
In other words, we sum over all possible insertion points of
each field  $\phi$ with a phase of the form $e^{ i 2\pi n j/J}$
where $j$ is the position of $\phi$ along 
the ``string of $Z$s''. We are interested in perturbative 
corrections to the dimension of the operator coming from the 
vertex \interterm . Since the $\phi$s are few and far apart
we can consider each insertion of $\phi$ independently, up
to $1/J$ corrections.

\ifig\mass{Diagrams that have ``momentum'', $n$, independent
contributions. }
{\epsfxsize 1 in\epsfbox{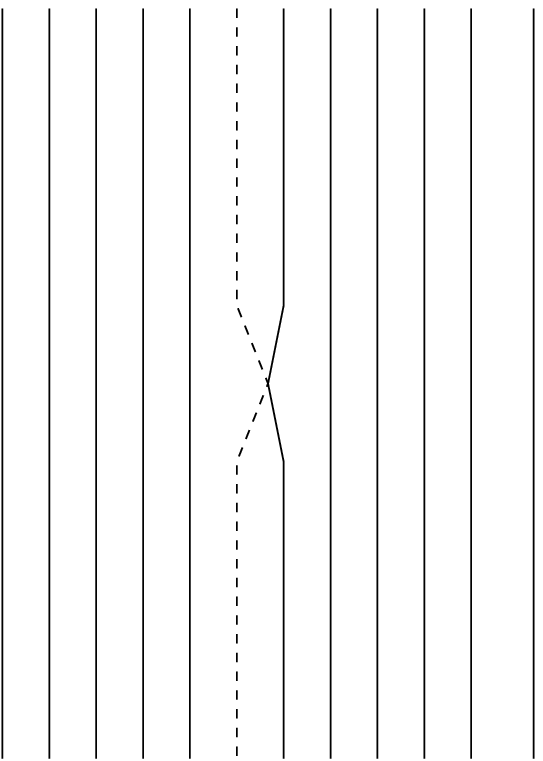}}

\ifig\massrest{Example of diagrams that have 
``momentum'' independent contributions that we do
not compute directly. These diagrams are the same
if we replace $\phi \to Z$. }
{\epsfxsize 1.7 in\epsfbox{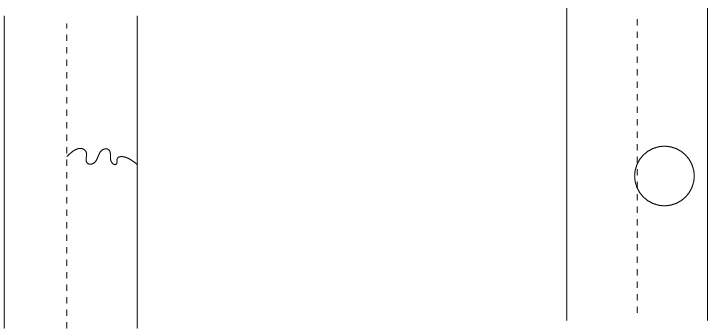}}

The terms in the (euclidean) 
Yang-Mills action that we will be interested
in are normalized as 
\eqn\normal{
S = { 1 \over 2 \pi g} \int d^4 x Tr\left({ 1 \over 2} (D\phi^I)^2 
- {1 \over 4} \sum_{IJ} [\phi^I,\phi^J]^2 \right)
}
where $I,J$ run over six values. We wrote the square of the 
 Yang-Mills coupling
in terms of what in $AdS$ is the string coupling that
transforms as $g \to 1/g$ under S-duality. The trace is just the 
usual trace of an   $N\times N$ matrix. 
We define $Z = {1 \over \sqrt{2}} ( \phi^5 + i \phi^6)$. Then the 
propagators are 
\eqn\prop{
\langle Z_{i}^{~j}(x) \bar Z_k^{~l}(0) \rangle 
= \langle \phi_{i}^{~j}(x) \bar \phi_k^{~l}(0)\rangle =
\delta_i^l \delta_k^j  \ {2 \pi g \over  4 \pi^2} \  { 1 \over  |x|^2}
}
In \normal\ there is an interaction term  of the form
the form ${ 1 \over { 2 \pi g}} 
\int d^4x   Tr( [Z,\phi][\bar Z , \phi])$, where
$\phi$ is one of the transverse scalars, let's say $\phi = \phi^1$. 
We  focus first  on the diagrams that give a contribution
that depends on the ``momentum'' $n$. These arise from interactions
that shift the position of $\phi$ in the operator, such as the ones
shown in  \exchange . These interactions come from a particular 
ordering of the commutator term in the action,  $ {1 \over 2 \pi g}
\int d^4 x 2 Tr[ \phi Z \phi \bar Z] $.  
These contributions give
\eqn\corr{
<O(x)O^*(0)> = { {\cal N} \over |x|^{2 \Delta} } \left[ 1+ N(2 \pi g)   
 4 \cos { 2 \pi n \over J}  I(x) \right]
}
where ${\cal N}$ is a normalization factor and  
 $I(x)$ is the integral 
\eqn\integr{
I(x) = {|x|^4 \over  (4 \pi^2)^2} \int d^4 y { 1 \over y^4 (x-y)^4 }
\sim { 1 \over  4 \pi^2} \log|x| \Lambda  + {\rm finite} 
}
We extracted the log divergent piece of the integral since it is the
one that
 reflects the change in the  conformal dimension of the operator. 

In addition to the diagrams we considered above 
 there are other diagrams, such as the ones  
 shown in  \mass\ and  \massrest,  which do not depend on $n$. 
We know that for $n=0$ the sum of all  diagrams 
cancels since in that case we have a protected operator and there is
no change in the conformal dimension. In other words, including the
$n$ independent diagrams amounts to replacing the cosine in \corr\
by 
\eqn\seno{
\cos{ 2\pi n \over J} -1  
}
In conclusion we find that for large $J$ and $N$  the first
correction to the $\phi$ contribution to the correlator is  
\eqn\corr{
<O(x)O^*(0)> = { {\cal N} \over |x|^{2\Delta} } \left[ 1 - 
{ 4 \pi g N n^2 \over J^2 } \log(|x|\Lambda) \right]
}
which implies that the contribution of the operator $\phi$ inserted 
in the ``string of $Z$s'' with momentum $n$ gives a contribution to
the anomalous dimension 
\eqn\contrib{
(\Delta - J)_n = 1 +  {2 \pi g N n^2 \over J^2}
}

There are similar computations we could do for insertions of 
$D_iZ$ or the fermions $\chi^a_{J=1/2}$. In the case of
the fermions the important interaction term will be
a Yukawa coupling of the form $ \bar \chi \Gamma_z [Z \chi] + 
\bar \chi \Gamma_{\bar z} [\bar Z , \chi] $. We have not done 
these computations explicitly since the 16  supersymmetries
preserved by the state \statezero\ 
relate them to the computation we did above for the insertion of
 a $\phi$ operator. 

The full square root arises from iterating these diagrams. 
This will be more transparent in the formalism we discuss 
in the next subsection.

\subsec{A Hamiltonian description}

In this subsection we reformulate the results of the previous 
subsection in a Hamiltonian formalism and we explain why 
we get a relativistic action on the string once we iterate the 
particular interaction that we are considering. 

Here we will consider the Yang-Mills theory defined on 
$S^3\times R$. All fields of the theory can be expanded in 
KK harmonics on $S^3$. States of this theory are in 
one to one correspondence with local operators on $R^4$. 
We take  the radius 
of $S^3$ equal to one so that 
 the energy of the state is equal to the conformal dimension
of the corresponding operator. 
For weak coupling, $g^2_{YM} N \ll 1$, 
the scalar fields give rise to a KK tower. The lowest energy 
state is the constant mode on $S^3$. Due to the curvature 
coupling there is effectively a mass term for the scalar fields 
with a mass equal to one (when the radius of $S^3$ is one). 
So the constant mode on $S^3$ is described by a harmonic oscillator
of frequency equal to one. Due to the color indices we have 
$N^2$ harmonic oscillators with commutation relations 
\eqn\commrel{
 [ a^{~i}_j , (a^\dagger)_k^{~l} ] = \delta_j^l \delta_k^i
}
for each mode. The fields $\phi^5$, $\phi^6$ lead to oscillators which 
can be combined into a pair of oscillators $a_+$ and $a_-$ with 
definite $J$ charge. From now one we denote by $a$ the $a_+$ oscillator. 
The operator  \statezero\ corresponds to the  state 
\eqn\statezeroosc{
{ 1 \over \sqrt{J} N^{J/2} } Tr[a^J] |0 \rangle
}
This is a single trace state. We will be interested  only in 
single trace states. In the large $N$ limit multiple trace states
are orthogonal to single trace states in the free 
theory\foot{This might not be true even in the free theory 
if $J$ is too large \refs{\vijay,\rango} 
but for our case where $J \sim N^{1/2}$ it is
indeed true. In the interacting theory we expect, from the gravity 
side,  non-planar corrections
when $J/\sqrt{N} \sim { 1 \over g_{YM}} $. }. 
In the free theory we can build all states by forming all possible 
``words'' out of all the oscillators associated to all the KK modes
of all the fields. The order is important up to cyclicity of the
trace. When we perform inner products or contractions of states
we will restrict only to planar contractions. Those are 
efficiently reproduced by replacing the standard oscillators 
$a_{\alpha \ j}^{~i}$, by Cuntz oscillators $a_\alpha$ where 
$\alpha$ labels the type of field and the KK mode \ref\voicu{
D. V. Voiculescu, K.J. Dykema and A. Nica  {\it Free Random
Variables} AMS, Providence (1992) } \ref\gg{R.~Gopakumar and D.~J.~Gross,
Nucl.\ Phys.\ B {\bf 451}, 379 (1995)
[arXiv:hep-th/9411021].
}.
The Cuntz algebra is 
\eqn\cuntz{
  a_\alpha a_\beta^\dagger = \delta_{\alpha,\beta}
}
and no other relation other than the one coming from the 
completeness relation 
\eqn\compl{
 \sum_\alpha a^\dagger_\alpha a_\alpha  = 1 - | 0 \rangle \langle 0|
}
More precisely, in order to take into account the factors of 
$N$ we replace $ { a_{\alpha \ j}^{~i} \over \sqrt{N}} \to a_\alpha $ where
the latter is a Cuntz oscillator. 
This algebra is rather useful for keeping track of the planarity 
of the contractions but one needs to be careful about 
enforcing properly the cyclicity of the trace, etc. 
As emphasized in \gg\ this algebra is a useful framework to 
study large $N$ matrix theories. 
In our case we will be interested in states of the form 
\eqn\states{
  \cdots a^\dagger b^\dagger a^\dagger\cdots a^\dagger b^\dagger 
a^\dagger \cdots |0\rangle
}
where the dots indicate a sequence of $a^\dagger$ operators. 
We will be interested in the action of the gauge theory 
Hamiltonian on such states where we have a small number of
$b^\dagger$. We will be interested in the 
interaction term in the Hamiltonian of the form 
\eqn\interham{
g^2_{YM} Tr( [Z,\phi][\bar Z \phi])  \to g^2_{YM} N  [a^\dagger,\phi][a,\phi]
}
where in the second term we think of $\phi \sim  b + b^\dagger$ where
$\phi$ is one of the transverse scalar fields and $b$ is the 
corresponding Cuntz oscillator. We neglect self contractions
in the Hamiltonian since those will be canceled by other 
propagator corrections in the case of ${\cal N} =4$ SYM. 
In the left hand side of the interaction term there are many possible
oscillators in the fields $Z$, $\bar Z$, we have only keep the piece
involving the oscillator with $\Delta-J=0$. 
An interaction  amounts  to an insertion of  the 
Hamiltonian \interham\ in any position of the state \states . 
We also need to sum in the right hand side of \interham\ over
 all possible orderings. 
Since there is a large number of $a^\dagger$ in the state \states\ we
an define $b_j$ oscillators which are the $b$ oscillators inserted 
at the $j$th position along the string. 
In this way the effective hamiltonian reduces to 
\eqn\effham{
H = \sum_j b^\dagger_j b_j  + 
{ g N \over 2 \pi } (b_j + b^\dagger_j - b_{j+1} -
b^\dagger_{j+1} )^2 
}
where again, as in the computation of the last subsection, the 
term proportional to $b^\dagger_i b_i$ was obtained indirectly 
by appealing 
to the BPS property of the state in the case of zero 
momentum excitations. 
We can see from this effective hamiltonian that the first 
correction on a state of the form 
$ \sum_l e^{i 2 \pi n l \over J} b_l^\dagger
|0\rangle'$ is indeed \firstcorr . Here the vacuum $|0\rangle'$ for
the $b_l$ oscillators is really the state \statezeroosc .
The effective hamiltonian is then essentially 
the discretized hamiltonian  of a massive scalar 
field in 1+1 dimension, 
where we discretize only the space direction, except that 
the oscillators in \effham\ have the usual commutation relations 
 for different sites but they are Cuntz oscillators on the same
site. 
We can see, however, that
if we define the oscillators
\eqn\fourmodes{
b_n^\dagger
 \equiv { 1 \over \sqrt{J}} \sum_{l=1}^J  e^{i 2 \pi l n \over J} 
b_l^\dagger
}
then the $b_n$ oscillators obey the standard commutation relations 
up to terms of order $1/J$ which we neglect in the large $J$ limit. 
For this reason the large $J$ limit of \effham\  will give 
the same as the continuum hamiltonian 
\eqn\contham{
H = \int_0^L d \sigma  \half \left[
\dot \phi^2 + \phi'^2 + \phi^2 \right] ~,~~~~L = J\sqrt{\pi \over  g N}
}

\ifig\randompath{ This is a schematic representation of the diagrams
that we are resumming to obtain \contham . The doted line is the 
$\phi$ propagator. Each crossing of a  $Z$ line is an interaction. 
}
{\epsfxsize .6in\epsfbox{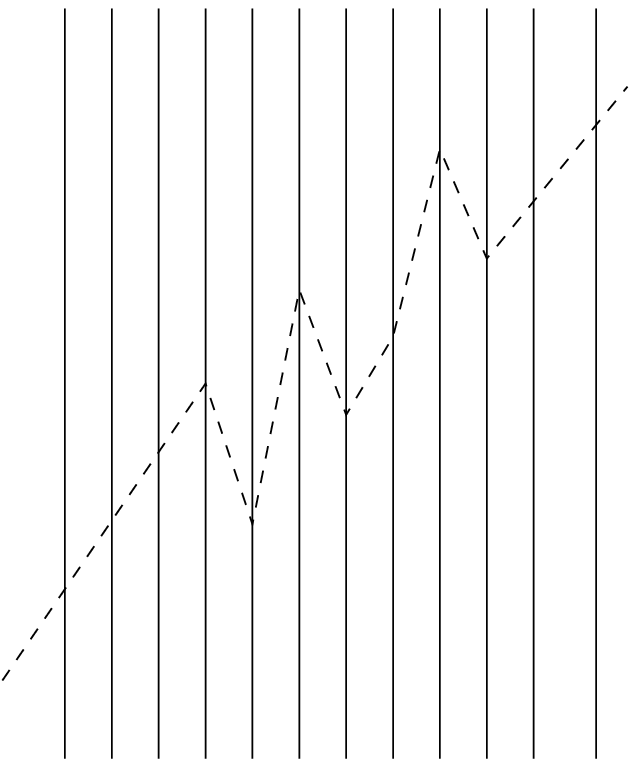}}

In \randompath\ we see see the form of the diagrams that we
are summing to obtain \contham . 
Note that when we diagonalize the new Hamiltonian \effham\ the
new vacuum will be related to the old vacuum by a Bogoliubov
transformation, so that in a sense there will be a fair number of 
$b^\dagger$s in the new vacuum. 
Supersymmetry ensures that the vacuum energy
does not change, so that we still have $\Delta-J=0$ for the new 
vacuum.

\subsec{ The fate of the other fields}

Let us now understand what happens when we insert in \statezero\ 
a field with $\Delta -J >1$. We will study the simplest case 
which arises when we insert the field  $\bar Z$. This field has
$\Delta-J = 2$. We can insert this field with arbitrary ``momentum'' $n$
in the operator (as long as we make sure that \momconstraint\ is obeyed). 
We will now show that the correction to its
dimension now does not vanish for zero momentum. 
We consider an operator of the form 
\eqn\operatorbar{
\sum_{l} e^{i 2 \pi l n \over J}  Tr[ \cdots ZZ \bar Z Z \cdots ]
}
where $l$ indicates the position of $\bar Z$ along the 
``string of $Z$s'' and the dots indicate a large number of $Z$ fields
together with possibly other insertions of other fields, etc. Since
we work in the dilute gas approximation, where $J$ is very large, we
can consider $\bar Z$ in isolation from other insertions of 
other operators. 
We can now compute the first order 
correction, in $gN$, to the 
anomalous dimension of \operatorbar . The relevant diagrams come
 from a vertex of the form $ { -1 \over 2 \pi g} \int d^4 x \half 
Tr( [Z,\bar Z])^2 $. The computation of these diagrams is identical 
to the one done in the first subsection of this appendix, the
only difference comes when we consider the combinatoric factors
in the diagram. There are again other diagrams (similar to those
in \massrest ) involving the 
exchange of gauge fields,  corrections to the propagator, etc, 
 which
we can effectively compute by noticing that if we change, for $n=0$,  
$\bar Z \to Z $ in \operatorbar\ , then we have a BPS state and all
diagrams should cancel. Putting this all together we obtain 
\eqn\correction{
 (\Delta- J)_n  = 2  +  { gN \over 4\pi} ( 4  +  {4 \pi^2 n^2 \over J^2})
+ \cdots 
}
where we expanded the result in powers of $1/J$. 
In contrast to \contrib\ we now find a contribution that is not
finite in the $N \to \infty$ limit that we are taking. 
We have computed
the correction only to first order in $gN$ and we are extrapolating
to $gN \to \infty$.  This is  not justified. 
So the above
computation should be taken as an indication that 
insertions of $\bar Z$ do not lead to finite energy excitations
in the effective 1+1 dimensional theory in the large $N$ limit
that we are taking. 

As we explained above, we expect that for $\Delta -J=1$ the fields
do not get a large mass because they are Goldstone bosons or
fermions of the symmetries broken by \statezero . 

\ifig\decayzbar{
(a) This diagram contributes to the decay of $\bar Z$ into two
$\phi$s. (b) This diagram is zero }
{\epsfxsize 1in\epsfbox{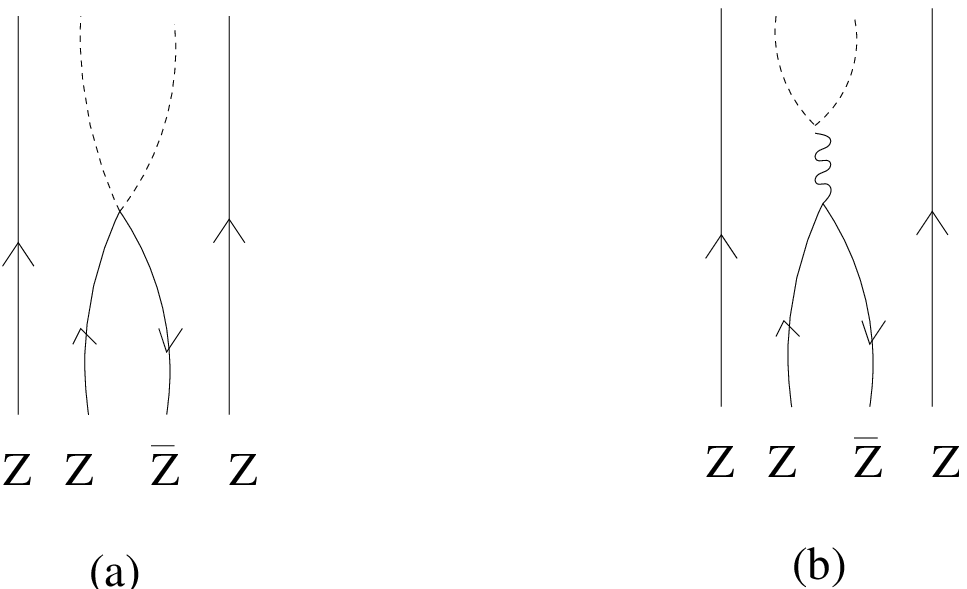}}

We can similarly compute the decay amplitude, to first order in 
$gN$, of the excitation with $\Delta - J=2$, created by the
insertion of $\bar Z$,  into excitations with $\Delta - J=1$ 
created by a pair of insertions of the transverse scalars
$\phi^r \phi^r$, $r =1,2,3,4$. These are given by the diagrams such
as the one  shown
in  \decayzbar (a). Again we find a result 
  proportional to $gN$ with no powers of $1/J^2$ to suppress it. 

In summary, we expect that in the large $N$ limit all excitations
created by fields with $\Delta -J >1$ become very massive and
rapidly decay to excitations with $\Delta - J=1$.

\appendix{ B}{Supersymmetry of the massive matrix model}

\subsec{Symmetry algebra}

In this subsection we make some remarks about the super 
 symmetry algebra. 
We will consider the 11d wave \fp\ but similar remarks apply for
the 10 dimensional waves \figueroaiib .

We define a generator $e = -p_-$ and a generator $h = -p_+$. 
The generator $e$ commutes with all the other operators. 
Some of the (anti)commutation relations are 
\eqn\comm{\eqalign{
[a_i, a_j^{\dagger}]= & e \delta_{ij} \;\;\;\;~~~~~~~i,j=1,..,9 \cr
[h, a_i^{\dagger}]= & {\mu \over 3}a_i^{\dagger}~,~~~~~~~
[h, a_i]=-{\mu \over 3}a_i \; \;\;\;~~ i=1,2,3 \cr
[h, a_i^{\dagger}]= &  {\mu \over 6} a_i^ {\dagger}~,~~~~~~
[h, a_i]=  -{\mu \over 6} a_i ~,~~~~~~i,j=4,..,9 
\cr 
\{ b_{\alpha \dot \beta }, b^{\dagger \ \gamma \dot \delta} \}= & e 
\delta_\alpha^\gamma \delta_{\dot \beta}^{\dot \delta}  
~~~~~0=\{ b, b' \} = \{ b^\dagger , {b'}^\dagger\}=[b,a]=
[b,a^\dagger]  \cr
[h,  b^{\dagger \ \gamma \dot \delta}] = &{ \mu \over 4 } 
b^{\dagger \ \gamma \dot \delta}
  ~,~~~~~~~
[h, b_{\alpha \dot \beta }] = - { \mu \over 4 } b_{\alpha \dot \beta } 
\cr
[h, Q_{\alpha \dot \beta}] =  & 
{ \mu \over 12}  Q_{\alpha \dot \beta} ~,~~~~~~~
[h, S^{\gamma \dot \delta}] = - { \mu \over 12} S^{\gamma \dot \delta}
\cr
\{ Q, Q\} = & \{S, S\} =0 ~,~~~~~~~~S^\dagger_{\alpha \dot \beta} =
Q_{\alpha \dot \beta}
\cr
\{ Q_{\alpha \dot \beta}, S^{\gamma \dot \delta}\} = &
\delta_\alpha^\gamma \delta_{\dot \beta}^{\dot \delta} h 
+ i {\mu \over 6 } 
\delta_{\dot \beta}^{\dot \delta} \sum_{i,j \leq 3} 
(\gamma_{ij})_{\alpha}^\gamma M_{ij} + 
i { \mu \over 12} 
\delta_\alpha^\gamma\sum_{i,j \geq 4} 
(\Gamma_{ij})_{\dot \beta}^{\dot \delta} M_{ij}
}}
where the undoted greek indices indicate spinor indices of SU(2)
an the doted ones denote spinor indices of SO(6) (the ones
downstairs are in the ${\bf 4}$ of SU(4) and the upstairs one are in the
${ \bf {\bar 4}}$ of SU(4)) and $\gamma^i$ and $\Gamma^j$ are three
and six dimensional gamma matrices respectively \foot{
The relation of the generators in \comm\ and those in \fp\ is
schematically as follows $ a^j \sim e^j + i {e^*}^j $ and similarly 
for $a^\dagger$, the $b$s and $b^\dagger$s are linear combinations
of $Q_+$ in \fp\ and similarly $S$ and $Q$ are linear combinations
of $Q_-$ in \fp .}.
In addition we have (anti)commutators of the $S$ and $Q$ with 
$b$s or $a$s which give $a$s or $b$s. We will not write those since
we will give them implicitly below when we discuss the superparticle. 
The main observation we want to make is that the structure of the 
representations of this algebra is very simple. Since $e$ commutes
with everything we can diagonalize it. Then the commutation 
relations of the $a$s and $b$s (and their adjoints) become 
bosonic and fermionic harmonic oscillators. Then the  rest 
of the symmetries acts  linearly on these oscillators. We can 
identify $h$ with the lightcone hamiltonian, so we see that
the $a^\dagger$ and $b^\dagger$ oscillators describe the 
center of mass motion of the state. In fact we could subtract
from $Q,S,h,M_{ij}$ an expression bilinear in these oscillators
(which is a realization of $Q,S,h,M_{ij}$ in terms of oscillators)
so than then $Q,S,h,M_{ij}$ act on the relative state. 
Note that $Q,S$ are supersymmetries that do not commute with the 
Hamiltonian. 

In the matrix model, the oscillators $a$ and $b$ are going to 
result from quantizing the $U(1)$ degree of freedom 
and the shift of $Q,S,h,M_{ij}$ that we mentioned above amounts to 
separating the $U(1)$ degree of freedom to leave the 
 $SU(N)$ degrees of
freedom.

\subsec{Plane wave limit of the 10d IIB $AdS_5 \times S_5$ 
action} 

Here we prove that the GS action of Metsaev \metsaev\ can be obtained 
as a limit of the $AdS_5 \times S_5$ action of \kt. 

There is a general formalism one can use in both cases.
Indeed, as shown in  \krr, for D branes propagating in 
supercoset manifolds, 
one can write down an action in terms of supervielbeins (vielbeins of 
the target superspace realized as a coset manifold). The kinetic term 
is always of the type 
\eqn\kin{
S=\int_M d^n\sigma \sqrt{g} g^{ij} L_i^{A}L_j^{A}
}
where $L_i^{A}$ are the bosonic components of the supervielbein
1-forms pulled back on the worldsheet. In general there can be also 
a WZ term, defined as the integral of a form on a n+1 dimensional manifold
with M as boundary.

The supervielbeins are found from the general procedure in \krr\
as 
\eqn\viel{
L^A= L_0^A +2 \theta ^{\alpha} f_{\alpha\beta}^A ({sinh ^2 M/2 
\over M^2})^{\beta}_{\gamma} (D\theta )^{\gamma}
}
and where the matrix M is defined by 
\eqn\tmatrix{
(M^2)^{\alpha}_{\beta}=-\theta^{\gamma} f_{\gamma A}^{\alpha} \theta^{
\delta}f_{\delta A}^{\beta}
}
the coefficients $f_{\alpha\beta}^A$ are the structure constants of 
the fermi-fermi part of the superalgebra $\{ F_{\alpha} , F_{\beta}
\} =f_{\alpha\beta}^A B_A$. If $\theta$ is constant, one gets the WZ
parametrization of superspace. Here 
\eqn\killing{
(D\theta)^{\alpha}= d\theta^{\alpha} + (L_0^A B_A\theta )^{\alpha}
}
is the Killing spinor operator acting on the $\theta$'s (the Killing 
spinor equation would be $D\epsilon (x)=0$).

The GS string action in a general supergravity background was given in 
\ghmnt\ and is
\eqn\gs{
S=-{1 \over 2} \int _{\partial M_3} d^2\sigma \sqrt{g} g^{ij} L_i^
{\hat{a}}L_j^{\hat{a}}+i\int _{M_3} s^{IJ} L^{\hat{a}}\wedge
\bar{L}^I\gamma^{\hat{a}}\wedge L^J
}
where $L^{\hat{a}}$ are the bosonic supervielbeins and $L^I$ the 
fermionic ones. 

In the case of $AdS_5 \times S_5$ the simple form of the action 
based on the above approach has been found in \ref\MetsaevIT{
R.~R.~Metsaev and A.~A.~Tseytlin,
Nucl.\ Phys.\ B {\bf 533}, 109 (1998)
[arXiv:hep-th/9805028].
}\kt:
\eqn\ads{
S=-{1 \over 2}\int d^2\sigma (\sqrt{-g} g^{ij}L^{\hat{a}}_i 
L^{\hat{a}}_j +4i\epsilon^{ij}\int_0^1 ds s^{IJ}L_{is}^{\hat{a}}
\bar{\Theta}^I \Gamma^{\hat{a}}L^J_{is})
}
where 
\eqn\adsviel{\eqalign{
L^I_s&= ({sinh(sM) \over M}D\Theta)^I\cr
L_s^{\hat{a}}&=e_{\hat{m}}^{\hat{a}}dX^{\hat{m}} -4i\bar{\Theta}^I
\Gamma^{\hat{a}}({sinh^2(sM/2) \over M^2}D\Theta)^I
}}
The fermionic light-cone gauge was fixed in \mt, and is the same 
as in flat space, namely $\Gamma^+\theta=0$. With this fermionic 
light-cone gauge, one gets that the matrix $M^2=0$, and so the 
only nontrivial information is encoded in $D\Theta$. But that has 
the general form 
\eqn\kill{
D\Theta^I=(\delta^{IJ}(d+{1 \over 4}\omega^{\mu\nu}\gamma^{\mu\nu}) 
+{i \over 48} e^{\mu}F_{\mu\mu_1...\mu_4}\Gamma^{\mu_1...\mu_4}
\epsilon^{IJ})\Theta^J
}
and consequently it has the correct limit from the $AdS_5\times 
S_5$ case to the pp wave case. The last step is the fixing of the 
bosonic light-cone gauge, which for the $AdS_5 \times S_5$ case 
was done in \mtt. Metsaev \metsaev, using the 
gauge
\eqn\gauge{
\sqrt{g}g^{ab}=\eta^{ab}\;\;\;\;\; x^+(\tau, \sigma)=\tau
}
finds then the action
\eqn\adsaction{
L=-{1 \over 2}\partial_ax^I \partial^a x^I -{\mu^2 \over 2} x_I^2 
-i\bar{\psi}\bar{\gamma}^-\rho^a\partial_a \psi +i\mu \bar{\psi}
\bar{\gamma}^-\Pi \psi
}

\subsec{Matrix theory action}

The action for  a single D0 brane  
can be obtained as the superparticle action moving in 
\metricelven\ in the Green-Schwarz
formulation. Indeed, for a D0 brane in flat space, the light-cone gauge 
superparticle action gives the free massless bosons $X^i$ and fermions 
$\theta$ (spinors of SO(9)), which is the free D0 action.

As we mentioned in the case of the GS string, the super-brane action has 
a kinetic and a WZ term. 
 But in the case of the superparticle, there is no 
2d form one can write down (except for the target space $AdS_2 \times S_2$
where one has the target space invariants $\epsilon_{ab}$). So the 
superparticle action has only the kinetic term.

The supervielbeins for the 11d supersymmetric pp-wave can 
be obtained as a limit from the $AdS_7 \times S_4$ supervielbeins, 
just as above for the 10d wave as a  limit of the $AdS_5 \times S_5
$. Indeed, from the above formalism, the supervielbeins can be written 
in a universal form depending only on the structure constants $f_{\alpha
\beta}^A$ of the superalgebra, and in terms of the Killing spinor 
operator. But we know that the wave space symmetry algebras are a 
contraction of the $AdS \times S$ ones, 
and that the Killing spinor operators
are also a similar limit (they only depend on F).

The supervielbeins for the $AdS_7 \times S_4 $ case have been given in 
\wpps. If one takes the general formulas there and substitutes
$F_{+123}=\mu$ and the fact that $\omega^{-i}$ are the only nonzero
components of $\omega^{\hat\mu\hat\nu}$one obtains
\eqn\killi{
D\theta= d\theta +{\mu \over 12}(e^{\hat{r}} {\Gamma_{\hat{r}}}
^{+123}-8e^{[+}\Gamma^{123]})\theta-{1 \over 2}
\omega^{-i}\Gamma_{-i}\theta
}
and also 
\eqn\msq{\eqalign{
(M^2)^{\alpha}_{\beta}&= {\mu \over 6} [({\Gamma_{\hat{r}}}^{+
123}-8\delta_{\hat{r}}^{[+}\Gamma^{123]})\theta]^{\alpha}(\bar{
\theta}\Gamma^{\hat{r}})_{\beta}\cr
&-{\mu \over 12} [(\Gamma_{\hat{r}\hat{s}}\theta)^{\alpha}(\bar{\theta
}\Gamma^{\hat{r}\hat{s} +123})_{\beta} +24 (\Gamma_{[-1}\theta)^{\alpha
}(\bar{\theta}\Gamma_{23]})_{\beta}]
}}
The superparticle action
\eqn\gen{
\int dt e^{-1} L_{t}^A L_{t}^A
}
will have a $k$ symmetry similar to the one of the free superparticle
with $L_t^{\mu}=\dot{x}^{\mu}-i\bar{\theta}^A\Gamma^{\mu} 
\dot{\theta}^A$. This $k$ symmetry needs to be 
gauge fixed by choosing the fermionic light-cone gauge. The procedure 
is exactly similar to the superstring in $AdS_5 \times S_5$ and its limit 
the 10 d wave (see \metsaev). As there, one can choose the 
gauge 
\eqn\gaugecd{
\Gamma^+\theta=0 \;\;\;\;\;\;\;\;\bar{\theta} \Gamma^+=0
}
which we can see from the expression of the $AdS_7 \times S_4$ $M^2$ above
that makes $M^2=0$, and so 
\eqn\waveviel{
L^A=dx^{\hat{\mu}}e_{\hat{\mu}}^A +{1 \over 2}
\bar{\theta}\Gamma^{A} D\theta
}
and where
\eqn\killin{
D\theta =d\theta +{\mu \over 12}(e^{\hat{r}}{\Gamma_{\hat{r}}}^{+123}
-8 e^{[+} \Gamma^{123]})\theta-
{1 \over 2}\omega^{-i}\Gamma_{-i}\theta
=d\theta +{\mu \over 12}e^+\Gamma^{-+}\Gamma^{123}\theta
-{\mu \over 6}e^+\Gamma^{123}\theta
}
where we have used the gauge condition to kill the terms with $\Gamma
^-$ and $\Gamma_{+i}$. 
One can then see that we get (in spacetime light cone parametrization)
\eqn\lmi{
L^+=e^+=dx^+ \;\;\;\;\;\;\;\; L^i=e^i=dx^i
}
and 
\eqn\lplus{
L^-= e^- + {1 \over 2}\bar{\theta}\Gamma^- D\theta ; \;\;\; 
e^{-}=dx^- -{1 \over 2}({\mu \over 3}
)^2 \sum _{i=1,2,3} (x^i)^2 dx^+ -{1 \over 2}
({\mu \over 6})^2 \sum_{i=4}^9(x^i)^2 dx^+
}
to be used in the action 
\eqn\acti{
S=\int dt (2L^+_tL^-_t + L^i_t L^i_t)
}
Then fixing the bosonic light cone gauge $ e=1, x^-(t)=t$ one gets the
action 
\eqn\matrixaction{
S=\int dt [(\dot{X}^i)^2 -({\mu \over 3})^2\sum_{i=1,2,3}(X^i)^2
-({\mu \over 6})^2\sum _{i=4}^9 (X^i)^2
 +\bar{\theta}\Gamma^- \dot{\theta} -{\mu \over 4}\bar{\theta}
\Gamma^-\Gamma^{123} \theta ]
}

We now rewrite the 11d fermions and gamma matrices in terms of 9d ones.
We choose the representation
\eqn\gammam{\eqalign{
\Gamma^{\mu}&=\gamma^{\mu}\otimes \sigma_3 \cr
\Gamma^0&=1\otimes i\sigma_2\cr
\Gamma^{11}&=1\otimes \sigma_1
}}

And we also choose a real (Majorana) representation for the spinors 
and gamma matrices: $C=\Gamma_0, \bar{\theta}=\theta^T C =\theta^{\dag}
\Gamma_0$. Then we have
\eqn\gammaid{
\Gamma^-=\sqrt{2} \pmatrix{ 0 & 0 \cr 1 & 0 \cr } 
\;\;\;\; \Gamma^+=\sqrt{2}  \pmatrix{ 0 & 1 \cr 0 & 0 \cr}
\;\;\;\; \Gamma_0\Gamma^-=-\sqrt{2} \pmatrix{
1 & 0 \cr 0 & 0 } \;\;\; \Gamma^{+-}=1\otimes \sigma_3
}

Then, take
\eqn\dimred{
\theta=\pmatrix{ &\psi_1 \cr &\psi_2 }
\;\;{\rm so} \;\; \Gamma^+\pmatrix{\psi_1 \cr \psi_2}
=0 \Rightarrow \psi_2=0
}

So, take
\eqn\dimredt{
\theta=\pmatrix{ \psi \cr 0 }
}
and so the fermion terms in the action sum up to
\eqn\fermion{
\sqrt{2}(\psi^T\dot{\psi}+{\mu \over 4}\psi^T\gamma^{123}\psi)
}
We can now absorb the $\sqrt{2}$ in front of this expression
in the definition of the fermions.

We turn to proving susy of this action, and generalizing it to the 
nonabelian case. We will leave the coefficient of the fermion mass term 
free, since we will find another solution for it in the abelian case.

Let us then start with the lagrangian 

\eqn\lagran{
L= \sum_{i=1}^9 (\dot{X} ^i)^2 -(\mu /3)^2 \sum_{i=1,2,3}
(X^i)^2 -(\mu /6)^2 \sum _{i=4}^9 (X^i)^2 +\Psi ^T \dot{\Psi}
-a(\mu /4)\Psi^T \gamma_{123} \Psi
}
and look for a susy transformation of the type
\eqn\susytr{\eqalign{
\delta X^i&=\Psi^T \gamma^i \epsilon (t) \cr
\delta \Psi &= \dot{X}^i\gamma^i \epsilon(t) +\mu X^i \gamma^i M'_i
\epsilon (t)\cr
\epsilon (t) &= e^{\mu M t} \epsilon_0
}}
Then the terms of order 1 in the susy transformation cancel , the 
terms of order $\mu$ give the equation 
\eqn\ordermu{
M'_i=\pm a/4 \gamma_{123} -M 
}
where the two values are for i=1,2,3 and i=4,..,9 respectively, 
and the terms of order $\mu^2$ give 
\eqn\ordermusq{\eqalign{ &
M'_i M -1/9 -a/4 \gamma_{123}M'_i=0 \;\;\; i=1,2,3\cr
& M'_i M -1/36 +a/4 \gamma_{123}M'_i=0\;\;\; i=4,..,9
}}
We then obtain 
\eqn\relations{\eqalign{
M&= b\gamma_{123}\cr
M'_i&=(\pm a/4 -b) \gamma_{123}
}}
and $a= 1$ or $ 1/3$ (2 solutions) and 
$b=-1/12$ or $- 1/4$ (the 2 corresponding solutions). There are also 
solutions where we change the sign of both a and b, but these correspond 
to the symmetry $\mu \rightarrow -\mu$.

The extension to the nonabelian theory is obvious; besides the 
usual commutator terms which are present in the lagrangian and 
susy rules in flat space, we have an extra coupling of order $\mu$.
Indeed, Myers \myers\ has found a term $F_{t ijk} Tr (X^i X^j X^k)$ 
in the action for N D0 branes in constant RR field. In our case, 
after the limit to the plane wave geometry (infinite boost),
the coupling is 
\eqn\coupling{
F_{+ ijk} Tr( X^i X^j X^k) \sim \mu Tr (X^i X^j X^k) \epsilon_{ijk}
}
So the lagrangian is 
\eqn\fullmatrix{\eqalign{
L&= \sum_{i=1}^9 (\dot{X} ^i)^2 -(\mu /3)^2 \sum_{i=1,2,3}
(X^i)^2 -(\mu /6)^2 \sum _{i=4}^9 (X^i)^2 +\Psi ^T \dot{\Psi}
-a(\mu /4)\Psi^T \gamma_{123} \Psi\cr &
+d \mu g \sum_{i,j,k=1}^3Tr (X^i X^j X^k) \epsilon_{ijk}
+2 g^2 Tr ([X^i, X^j]^2) +2 ig Tr (\Psi^T \gamma^i [\Psi, X^i])
}}

And the susy rules are 

\eqn\susyrule{\eqalign{
\delta X^i&=\Psi^T \gamma^i \epsilon (t) \cr
\delta \Psi &= \left( \dot{X}^i\gamma^i  +\mu X^i \gamma^i (\pm
a/4-b)\gamma_{123}+ ig [X^i, X^j] \gamma_{ij} \right) \epsilon(t) \cr
\epsilon (t) &= e^{\mu M t} \epsilon_0
}}

The terms of order $g^0$ in the susy transformation of L 
work the same way as for one D0 brane, since they are bilinear in fields.
The terms of order g cancel (they would fix the coefficient of the 
$\Psi \psi X$ term in the action). The terms of order $\mu g$
are proportional to $ Tr (\Psi^T \gamma^{ij} \gamma_{123} \epsilon (t) 
[X^i, X^j] )$ and split into i,j both =4,..,9, one of i,j =1,2,3 and 
the other =4,..,9 which both give the equation 
\eqn\baf{
3b +a/4=0
}
and the case when both i,j are 1,2,3 which gives
\eqn\dtg{ 
d=2 (b-a/4)
}
So now a and b are restricted to just $a=1, b=-1/12$.
This solution is the one we found from the general formalism.
The terms of order $g^2$ cancel (they would fix the coefficient of the 
$[X,X]^2$ term in the action). 

The action has the almost the same nonlinearly realized susy as 
in flat space. In flat space, the nonlinear susy is $\delta \Psi=
\epsilon$ (constant), and the X's constant. In our case, 
we have 
\eqn\nonlinsusy{
\delta \Psi =\epsilon (t) =e^{\mu {a \over 4} \gamma_{123} t} \epsilon_0
}

\appendix{C}{ Strings on mixed NS and RR plane waves}

As we remarked above, we can consider the limit of 
section 2, for the $AdS_3 \times S^3$ backgrounds. 
It is interesting to consider such a limit in a situation 
where we have a mixture of NS and RR three form field strength. 
The six dimensional  plane wave  metric that we obtain has the form  
\eqn\metric{\eqalign{
ds^2 = & - 4 dx^+ dx^-  - \mu^2  \vec y^{\ 2} {(dx^{+})}^2 + d\vec y^{ \ 2}
\cr
H^{NS}_{+12} &= H^{NS}_{+34} =  C_1 \mu \cos \alpha  \cr
H^{RR}_{+12} &= H^{RR}_{+34} = C_2   \mu \sin \alpha 
}}
where $\vec y$ parametrizes a point on $R^4$ and
 $\alpha$ is a fixed parameter which allows us to 
interpolate between the purely NS background $\alpha=0$ and
the purely RR background $\alpha = \pi/2$.
The constants $C_1, C_2$ depend on the string coupling and the 
normalization of the RR and NS field strengths.  
In addition to the six coordinates in \metric\ we
have four additional directions which we can take to be a $T^4$
(or a K3). 

The light cone action becomes 
 \eqn\lcactsix{\eqalign{
S =& {1 \over 2 \pi \alpha'} 
\int dt \int_0^{ 2\pi  \alpha' p^+} d \sigma \half
\left[  |\dot Z_i|^2 - 
|Z'_i + i \cos\alpha Z_i|^2 -  \sin^2 \alpha\mu^2 |Z_i|^2 \right] 
 +
\cr
&  \bar S ( \sigma^0\partial_0 + 
\sigma^1 (\partial_1 +  \cos\alpha I )  + \sin \alpha \mu I ) S 
+ {\cal L}_{T^4} 
}}
Where $x$ denotes the four coordinates of $T^4$ and 
$  I \equiv \Gamma^{12}$. 
We have also defined $Z_1= y_1 + i y_2$ and $Z_2 = y_3 + i y_4$. 
The fermions $S$ in \lcactsix\ have positive chirality 
in the directions 1234 (and hence also positive chirality 
in the $T^4$  directions). The lagrangian ${\cal L}_{T^4}$ includes 
 the modes living on $T^4$ as
well as for the fermions that have negative chirality 
on the directions $1234$ which are still massless. 
Only half of the fermions get a mass.

The light cone Hamiltonian is then 
\eqn\lchamsix{\eqalign{
p^+  =  H_{lc} = &\sum_{n=-\infty}^{\infty} 
N_n \sqrt{ \sin^2 \alpha^2 \mu^2 +  ( \cos  \alpha \mu + { n \over
  \alpha' p^+})^2}  ~~ +
\cr & +  2 {  L_0^{T^4} + \bar L_0^{T^4} \over  \alpha' p_+} 
}}
where the first line takes into account the massive bosons and
fermions and the second line takes into account the $T^4$ bosons
and the four massless fermions. We also have the condition that
the total momentum along the string is zero. 

We see that in the pure RR case we get something quite similar
to the previous result. For the pure NS case the spectrum 
can be viewed as arising from twisted boundary conditions 
along the string. In that case, when 
$  \alpha' p^+ \mu =  n$ we have a new zero mode appearing. 
When we excite this zero mode we obtain a string that winds
$n$ times around the origin and has zero light cone energy due
to the cancellation of the gravitational and  ``electric'' energy. 
These are analogous to the long strings much discussed in 
$AdS_3$ with NS background \refs{\longstrings}. 
As soon as $\cos\alpha \not = 1$ these new 
zero modes disappear, as is expected. 

It would be nice to see if it is possible to reproduce the 
spectrum \lchamsix\ from the dual CFT of the D1-D5 system.

\listrefs

\bye